\documentclass[11pt,a4paper]{article}
\pdfoutput=1

\usepackage{style}
\usepackage{rotating}
\usepackage{lscape}
\usepackage{multirow}
\usepackage{color,soul}

\usepackage[utf8]{inputenc}
\usepackage[T1]{fontenc}
\usepackage{mathrsfs}
\usepackage{amsmath}
\usepackage{amssymb}
\usepackage{amsthm}
\usepackage{psfrag}
\usepackage{graphicx}
\usepackage{hyperref}
\usepackage{color}
\usepackage{bm}
\usepackage{url}
\usepackage{graphicx}
\usepackage{cleveref}
\usepackage{physics}

\usepackage{bbm}
\usepackage{scalerel}
\usepackage{upgreek}

\usepackage[font=small,labelfont=bf,tableposition=top]{caption}
\DeclareCaptionLabelFormat{andtable}{#1~#2  \&  \tablename~\thetable}

\newcommand{\be}{\begin{equation}}
\newcommand{\ee}{\end{equation}}
\newcommand{\ba}{\begin{align*}}
\newcommand{\ea}{\end{align*}}
\newcommand{\beqa}{\begin{eqnarray}}
\newcommand{\eeqa}{\end{eqnarray}}
\newcommand{\bseq}{\begin{subequations}}
\newcommand{\eseq}{\end{subequations}}
\newcommand*\diff{\mathrm{d}}

\let\limitint\int 
\renewcommand{\int}{\limitint \!} 

\newcommand{\D}{\mathcal{D}}

\renewcommand\d{\partial}
\newcommand\G{\Gamma}
\newcommand\e{\text{e}}
\renewcommand\l{\lambda}
\newcommand\vf{\varphi}
\renewcommand\a{\alpha}

\renewcommand\Im{\text{Im}}
\newcommand\const{\text{const}}

\renewcommand\o{\omega}
\newcommand{\s}{\sigma}
\renewcommand{\O}{\Omega}

\newcommand{\TT}{\hat{T}}
\newcommand\Teff{T_{\rm eff}}
\newcommand\Es{\hat{E}_b}
\newcommand\heta{\hat{\eta}}

\newcommand{\sh}{\mathop{\rm sh}\nolimits}
\newcommand{\ch}{\mathop{\rm ch}\nolimits}
\renewcommand{\th}{\mathop{\rm th}\nolimits}

\usepackage[toc,page]{appendix}

\newcommand{\nocontentsline}[3]{}
\newcommand{\tocless}[2]{\bgroup\let\addcontentsline=\nocontentsline#1{#2}\egroup}

\relax

\title{
Thermal false vacuum decay in (1+1)-dimensions: Evidence for non-equilibrium dynamics
}

\author[a,b]{Dalila P\^irvu\footnote{\texttt{dpirvu@perimeterinstitute.ca}}}
\author[a]{Andrey Shkerin\footnote{\texttt{ashkerin@perimeterinstitute.ca}}}
\author[a,c]{Sergey Sibiryakov\footnote{\texttt{ssibiryakov@perimeterinstitute.ca}}}

\affiliation[a]{Perimeter Institute for Theoretical Physics, 31 Caroline St N, Waterloo, ON N2L 2Y5, Canada}
\affiliation[b]{Department of Physics \& Astronomy, University of Waterloo, Waterloo, ON N2L 3G1, Canada}
\affiliation[c]{Department of Physics \& Astronomy, McMaster University, 1280 Main Street West, Hamilton, ON L8S 4M1, Canada}

\abstract{
We numerically study the evolution of a classical real scalar field in ${(1+1)}$ dimensions with initial conditions describing thermal fluctuations around a metastable vacuum. We track false vacuum decay in real time and compare several observables to the predictions of the standard Euclidean formalism. We find agreement for the shape of the critical bubble and the exponential suppression of the decay rate. However, the decay rate prefactor is almost an order of magnitude lower than the predicted value. We argue that this signals a breakdown of thermal equilibrium during the bubble nucleation. In addition, the inefficient thermalization in the system biases the properties of the statistical ensemble and leads to further decrease of the decay rate with time.
We substantiate our interpretation with a suite of stochastic field simulations with controlled thermalization time. Varying this time we find that the predictions of the standard equilibrium formalism are recovered when it is sufficiently short. We propose an upper bound on the thermalization time that must be satisfied in order to ensure the applicability of the Euclidean rate calculation. We discuss that this bound is unavoidably violated in common single-field models, irrespective of the number of spacetime dimensions, implying that deviations from equilibrium in these models cannot be neglected. In theories with multiple fields, the bound may or may not hold, depending on the setup details.
We investigate one more signature of non-equilibrium dynamics --- coherent oscillonic precursors to the critical bubble nucleation. We show that they get suppressed in the stochastic dynamical simulations when the thermalization time is reduced.

\begin{flushright}
{\it Dedicated to the memory of Valery Rubakov}
\end{flushright}

}

\begin{document}

\maketitle

\section{Introduction and summary}
\label{sec:intro}

Theoretical studies of false vacuum decay have a long history. They can be traced back to Gibbs' analysis of first-order phase transitions through nucleation of critical bubbles of the new phase inside a metastable phase~\cite{gibbs1928collected}. The similar problem of a metastable state activation arising in physical chemistry was addressed by Wigner~\cite{wigner1938transition} and Kramers~\cite{KRAMERS1940284}, whose works laid the foundations of the modern reaction rate theory, see~\cite{Hanggi:1990zz} for review. 
This approach was generalized by Langer to thermal activation in classical systems with multiple degrees of freedom~\cite{Langer:1969bc}. Finally, a full quantum treatment of false vacuum decay in field theory was developed for zero~\cite{Kobzarev:1974cp, Coleman:1977py, Callan:1977pt} and finite~\cite{Linde:1980tt, Linde:1981zj, Affleck:1980ac} temperature.

First-order phase transitions could have occurred in the early universe~\cite{Mazumdar:2018dfl}. They have been widely discussed in the context of inflation~\cite{Guth:1980zm, Guth:1982pn, Turner:1992tz}, baryogenesis~\cite{Rubakov:1996vz, Bodeker:2020ghk} (together with the related sphaleron processes~\cite{Kuzmin:1985mm}) and as a promising source of primordial gravitational wave background~\cite{Caprini:2018mtu, Caprini:2019egz, Athron:2023xlk}. Fully uncovering their phenomenology requires accurate predictions for various observables characterizing their dynamics, such as the transition rate, size and shape of the bubbles of new phase, their expansion rate, etc.

Another motivation for refining the theoretical description of false vacuum decay comes from recent advances in manipulation of atomic systems. These have led to proposals for realizing false vacuum decay in the laboratory using quantum simulators~\cite{Fialko:2014xba, Fialko:2016ggg, Braden:2017add, Billam:2018pvp, Braden:2019vsw, Billam:2021nbc, Jenkins:2023eez, Jenkins:2023npg}, and the first successful experiment has been reported in~\cite{Zenesini:2023afv}. This opens the way to experimental tests of the false vacuum decay dynamics in real time.

The traditional approach to false vacuum decay is based on the Euclidean time formalism (see e.g.~\cite{Weinberg:2012pjx}). It treats the bubble nucleation rate $\Gamma$ as a thermodynamic quantity and relates it to the imaginary part of the false vacuum free energy. At temperature $T$ sufficiently above the mass of the decaying field $m$, this relation reads~\cite{Affleck:1980ac},\footnote{We use units $\hbar=c=1$ and define the nucleation rate as the number of bubbles forming per unit time and volume.}
\be
\label{GammaAff}
\Gamma_E=\frac{\omega_-}{\pi T}\cdot \frac{\Im F}{\cal V}\;,
\ee
where $\omega_-\sim m$ is the critical bubble growth rate, and ${\cal V}$ is the volume of the system. The critical bubble arises in this approach as the static solution of the equations of motion corresponding to the saddle point of the path integral, which defines the system's partition function. 
The imaginary part of the free energy $\Im F$ is evaluated in the Gaussian approximation around this saddle point. The contribution of the critical bubble into the free energy is exponentially suppressed by the Euclidean bubble action and it is imaginary because of the presence of an unstable mode 
with the eigenvalue~$-\omega_-^2$.  

This picture raises several questions. One of them concerns the determination of the bubble profile and its action. Shall one use for this purpose the bare energy functional of the system, or the free energy which includes contributions from quantum and thermal fluctuations?
This question is particularly pertinent since there are many 
systems where the very existence of metastable phase relies on thermal corrections to the effective potential. The answer is by now reasonably well understood. The critical bubble at high temperature is a saddle point of the effective free energy $ {\cal F}$  containing contributions of all fluctuations, {\it except} the zero Matsubara mode of the decaying field itself~\cite{Gleiser:1993hf, Bodeker:1993kj, Gould:2021ccf} (see also~\cite{Weinberg:1992ds} for the zero-temperature case). 
While this prescription allows us to calculate the right hand side of eq.~(\ref{GammaAff}), it tells us little about how the relevant bubbles emerge in real time, or even in what sense their profiles are expected to match the theoretical curve, given that all fields are subject to rapid thermal fluctuations.

Another question is even more fundamental and concerns an assumption deeply built in the Euclidean formalism that the nucleation of bubbles is an equilibrium process. This is far from clear since the bubble nucleation is precisely the process that drives the system to a new phase, i.e. far away from the original equilibrium state. Moreover, the thermalization time $t_{th}$ in a weakly coupled theory is typically much longer than the dynamical time scale of the bubble set by $\omega_-^{-1}\sim m^{-1}$, suggesting an inevitable breakdown of equilibrium during the nucleation. We are going to see that this reasoning is somewhat too naive and the comparison between $t_{th}$ and $\omega_-^{-1}$ should involve an additional factor. Nevertheless, it correctly points out the problem.

A different derivation of the thermal false vacuum decay rate, not relying on the Euclidean formalism, is due to Langer~\cite{Langer:1969bc}. He considered transitions over a potential barrier in a classical-statistical system driven by the Langevin equation with the friction and noise related by the fluctuation-dissipation theorem. They represent the interaction of the system with the heat bath, and the friction coefficient $\eta$ sets the thermalization time, $t_{th}\sim \eta^{-1}$. The expression for the rate then reads,
\be
\label{GammaLan}
\Gamma_{\rm stat}=\frac{1}{\pi T}\left(\sqrt{\omega_-^2+\frac{\eta^2}{4}}-\frac{\eta}{2}\right)\cdot \frac{\Im F}{\cal V}\;.
\ee
It differs from eq.~(\ref{GammaAff}) by the dynamical prefactor, and reduces to it in the limit $\eta\to 0$. However, as we review in Appendix~\ref{app:Langer}, this classical-statistical derivation is only valid at strong enough friction,
\be
\label{etaLan}
\eta\gg  \frac{\omega_- T}{{\cal F}_b}\;,
\ee
where ${\cal F}_b$ is the effective free energy of the critical bubble. Thus, the extrapolation of eq.~(\ref{GammaLan}) to zero friction is not justified. The condition (\ref{etaLan}) is well known in the reaction rate theory studying transitions in mechanical systems with just a few degrees of freedom. For such systems, the expression (\ref{GammaLan}) breaks down once (\ref{etaLan}) is violated, and in fact, the classical transition rate vanishes at $\eta=0$~\cite{Hanggi:1990zz}. However, we are not aware of any discussion of the condition (\ref{etaLan}) in the field theory context.

Yet another approach to non-perturbative transitions at high temperature was pioneered by Grigoriev and Rubakov~\cite{Grigoriev:1988bd}. It makes use of the observation that at high $T$ the occupation numbers of the long wavelength modes responsible for the transition are large, so the fields can be treated as classical. Then, upon randomly picking initial conditions from an ensemble of thermal fluctuations around the original vacuum, one can numerically evolve the system on the lattice and trace the transitions in real time. This approach has been applied to sphaleron transitions~\cite{Grigoriev:1989je, Grigoriev:1989ub, Ambjorn:1990wn, Ambjorn:1990pu}, production of topological solitons~\cite{Alford:1991qg, Bochkarev:1989tk, Habib:1999mc}, and false vacuum decay~\cite{PhysRevB.42.6614, Alford:1993zf, Alford:1993ph, Borsanyi:2000ua, Gleiser:2004iy, Gleiser:2007ts, Pirvu:2021roq, Batini:2023zpi, Pirvu:2023plk}. 
It has proved fruitful in uncovering the features of the real-time dynamics of the phase transitions which are left outside the scope of the Euclidean method, such as non-zero bubble velocities at nucleation and oscillonic bubble precursors~\cite{Gleiser:1991rf, Gleiser:1993pt, Gleiser:2004iy, Gleiser:2007ts, Aguirre:2011ac, Pirvu:2023plk}. Hybrid versions of this method that combine the Euclidean Monte-Carlo sampling with real-time evolution of configurations with bubbles were also developed~\cite{Moore:2000jw, Moore:2001vf, Gould:2022ran, Gould:2024chm} to address the case of strongly suppressed phase transitions when a direct numerical evolution from the metastable state is unfeasible. 

We follow the real-time approach in the present work in order to perform precision tests of the Euclidean theory predictions for the bubble profile and the nucleation rate. To make the comparison as clean as possible, we restrict to a single scalar field in $(1+1)$ dimensions with negative quartic potential. In this setup the bubble profile is expected to follow from the bare potential.\footnote{Note that the hard thermal loops that introduce large thermal mass renormalization in $(3+1)$ dimensional theories~\cite{Parwani:1991gq, Aarts:1997kp} are absent in $(1+1)$d. The only mass renormalization then comes from soft classical modes~\cite{Boyanovsky:2003tc}.}  
The Euclidean prediction for the rate then takes the form, 
\be 
\label{G}
    \Gamma_E = A_E\,\e^{-E_b/T} \;,
\ee
where $E_b$ is the critical bubble energy and the prefactor $A_E$ is also calculable, see eq.~\eqref{G_E2} below.

We prepare a suite of real-time numerical simulations starting from random thermal initial conditions around the metastable vacuum and trace their evolution until decay. Counting the number of surviving systems as a function of time allows us to extract the decay rate, while the inspection of the field configurations at times close to the decay provides information about the properties of the critical bubble and dynamics of its nucleation. We perform several simulation runs exploring both the Hamiltonian (i.e. conservative) and Langevin form of the dynamics. The Langevin case models the coupling of the system to an external heat bath and allows us to vary the thermalization time of the system, $t_{th}\sim \eta^{-1}$, by changing the dissipation parameter $\eta$. Our main results are summarized as follows:
\begin{itemize}
\item We find that the critical bubble profile obtained by averaging over multiple stacked configurations indeed reproduces the static solution of the bare equations of motion, in agreement with the Euclidean theory. We do not detect any thermal corrections to it. This is non-trivial since thermal corrections to the scalar mass are sizable for the parameters of our simulations and affect the preparation of the thermal ensemble.
\item The decay rate measured in the Hamiltonian simulations is lower than the Euclidean prediction by almost an order of magnitude. Furthermore, it slowly decreases with time. The discrepancy is due to inefficient thermalization and signals a breakdown of the Euclidean approach.
\item For the Langevin simulations the decay rate depends on the value of the friction parameter $\eta$:
\begin{itemize}
\item For 
$
\eta\gtrsim \omega_-T/E_b
$
the rate is constant in time and consistent with the Langer's result \eqref{GammaLan}, which can be obtained from eq.~(\ref{G}) by rescaling the prefactor, see \cref{pref_Langer} below.
\item For $\omega_-T/E_b \gtrsim \eta\gtrsim t_{dec}^{-1}$
the rate is still constant, but deviates significantly from eq.~(\ref{GammaLan}). Here $t_{dec}$ is the average decay time of a single system, $t_{dec}\sim (\Gamma L)^{-1}$, with $L$ being the length of the simulation box.
\item For $t_{dec}^{-1}\gtrsim \eta$
the rate follows the same time dependence as for the Hamiltonian simulations.
\end{itemize}
\item We propose that for a general system the applicability of the equilibrium formulae (\ref{GammaAff}), (\ref{GammaLan}) requires an upper bound on the thermalization time,\footnote{In this criterion we assume that the temperature is high, $T\gg \omega_-$, so that the transition is classical.}
\be \label{criterion}
    t_{th}\lesssim\frac{{\cal F}_b}{\omega_-T}\;.
\ee
Since the ratio ${\cal F}_b/T$ sets the exponential suppression of the decay rate, it must be bigger than unity. Hence the bound (\ref{criterion}) is weaker 
than the naive requirement 
${t_{th}\lesssim \omega_-^{-1}}$ which would follow from the mere comparison of the thermalization and bubble growth rates. Nevertheless, eq.~(\ref{criterion}) is always violated in the Hamiltonian dynamics of a single scalar field, irrespective of the number of dimensions. For theories with multiple fields, it may or may not be satisfied, depending on the details of the interactions.  
\item We study another manifestation of the non-equilibrium dynamics accompanying the bubble nucleation --- the oscillonic precursors~\cite{Pirvu:2023plk}. These are coherent localized oscillations of the field preceding the bubble formation. We find that they get suppressed in the Langevin simulations and disappear completely at strong friction. This is consistent with the expectation that coupling to a heat bath erases non-equilibrium features.
\end{itemize}

The above results were partially presented in~\cite{Pirvu:2024ova}. In the rest of this paper we elaborate on their derivation and give details of our analytical and numerical methods.

The paper is organized as follows.
In Sec.~\ref{sec:setup} we describe our setup.
Following the standard formalism, we derive the critical bubble solution and the Euclidean decay rate, including the prefactor. 
In Sec.~\ref{sec:num} we introduce the discretized version of the theory, discuss its thermal properties and estimate, both analytically and numerically, its thermalization time. Section~\ref{sec:sim} is dedicated to the Hamiltonian simulations of false vacuum decay. We recover the critical bubble profile, describe the measurement of the decay rate and perform a detailed comparison with the Euclidean prediction. 
In Sec.~\ref{sec:simL} we model the coupling of the theory to an external heat bath by supplementing its equation of motion with dissipation and noise terms. We study how increasing the dissipation and noise modifies the dynamics of the phase transition. 
In Sec.~\ref{sec:criterion} we formulate the criterion for the applicability of the standard equilibrium nucleation picture and point out that it is, in general, violated in commonly studied field theories.
In Sec.~\ref{sec:osc} we study oscillons preceding the formation of the critical bubble and show that such coherent long-lived structures can only live in a system with weak dissipation and noise. Section~\ref{sec:concl} contains discussion and outlook. Several appendices contain technical details and auxiliary derivations.

\section{Setup}
\label{sec:setup}

\subsection{Action and the classical limit}
\label{ssec:effpot}

We consider the theory of a real scalar field in $(1+1)$ dimensions with the action in the $(-+)$ signature,
\be \label{S}
    S = \int\diff t \, \diff x\left( -\frac{(\d_\mu\phi)^2}{2} -V(\phi) \right) \;,
    \qquad \qquad V(\phi)=\frac{m^2\phi^2}{2} - \frac{\l \phi^4}{4}\;,
\ee
where $\l>0$.
The false vacuum is located at $\phi=0$, and the true vacuum corresponds to the run-away $\phi\to\pm\infty$. The system is at finite temperature $T$.
The partition function of the system reads,
\be \label{Zgen}
    Z  = \int [\diff\phi]\:\e^{-S_E [\phi] } \;,
\ee
where the integral runs over Euclidean configurations with period $1/T$ in the Euclidean time. 

Before proceeding to the false vacuum decay it is instructive to clarify the relation between the full quantum theory and its classical high-temperature limit. To this end, we compute the 1-loop effective potential. Expanding the field around a constant background $\bar{\phi}$ and integrating over the fluctuations $\delta \phi$, we obtain  
\be
\label{V-1-loop-int}
V_{\rm eff}(\bar{\phi})\equiv -\frac{T}{L}\ln Z=V(\bar{\phi})
+\frac{T}{2}\sum_{n=-\infty}^\infty\limitint_{-\infty}^\infty \frac{dp}{2\pi}
\ln\left( 1+ \frac{m^2(\bar{\phi})}{\omega_n^2+p^2} \right)\;,
\ee
where in the second equality we have subtracted a $\bar{\phi}$-independent divergent constant. Here $L$ is the linear size of the system, $\omega_n=2\pi Tn$ are the Matsubara frequencies, and $m^2(\bar{\phi}) = m^2-3\l\bar{\phi}^2$ is the background-dependent mass of the fluctuations. The sum over the Matsubara frequencies diverges. In Appendix~\ref{app:det} we show that the divergence is absorbed by renormalizing the mass at $T=0$. Then the contribution of the sectors with $n\neq 0$ becomes negligible in the classical limit $T\gg m$, and the effective potential is saturated by the contribution of the zero Matsubara mode,
\be \label{V-1-loop-reg}
    V_{\rm eff}(\bar{\phi}) \simeq V(\bar{\phi})+ 
  \frac{T}{2}\sqrt{m^2-3\l\bar{\phi}^2} \;.
\ee
Note that this situation is different from that in $(3+1)$ dimensions where $n\neq 0$ sectors produce large corrections to the field mass, $\delta m^2\sim \l T^2$, which must be incorporated in the connection between the parameters of the classical and quantum theories, see e.g.~\cite{Parwani:1991gq, Aarts:1997kp}.

Expression (\ref{V-1-loop-reg}) allows us to make two observations. First, it implies a thermal correction to the mass of the field in the metastable minimum,
\be \label{mth}
    m^2_{\rm th} = m^2 - \frac{3 \l T }{2m} \;.
\ee
Since the potential (\ref{V-1-loop-reg}) is entirely produced by the zero Matsubara mode, this mass shift is a purely classical effect. Its size relative to the bare mass is characterized by a dimensionless temperature parameter
\be \label{NewTemp} 
    \TT = \frac{\l T}{m^3} \;,
\ee
which controls the strength of classical interactions in the model, cf.~\cite{Boyanovsky:2003tc}.\footnote{Note that if we restore $\hbar$ in the definition of various quantities, it will cancel out in $\TT$. } The theory remains weakly coupled as long as 
\be \label{boundT}
    \TT \ll 1 \;.
\ee
Second, the effective potential (\ref{V-1-loop-reg}) is real only for 
$\bar{\phi}<m/\sqrt{3\lambda}$. 
This is the inflection point of the tree-level potential, beyond which the fluctuations become tachyonic. 
Hence, for large deviations from the metastable vacuum, such as the critical bubble discussed in the next subsection, the effective potential is ill-defined.\footnote{Some works use the real part of the effective potential when evaluating the effect of thermal corrections on vacuum decay. This prescription, however, is not justified, see e.g. the discussion in \cite{Croon:2020cgk}.}

\subsection{Critical bubble and the Euclidean decay rate}

\begin{figure}[t]
    \centering
    \includegraphics[width=0.55\linewidth]{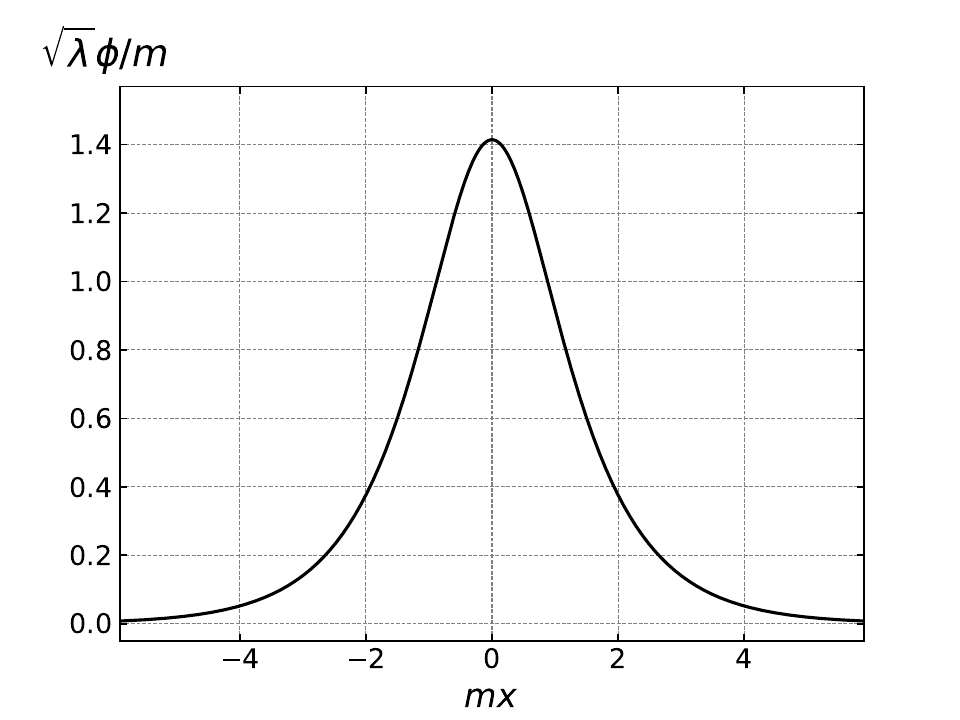}
    \caption{Critical bubble in the theory (\ref{S}).}
	\label{fig:sph_th}
\end{figure}

We are interested in the decay of the metastable vacuum $\phi=0$ 
 via classical thermal jumps of the field over the barrier.
Following the standard approach, we look for the saddle point of the partition function (\ref{Zgen}). The relevant solution is the static critical bubble.  
It satisfies the equation
\be \label{spheq1}
    -\phi_b^{\prime\prime} + m^2\phi_b - \l \phi_b^3 = 0 \; .
\ee
Imposing vanishing boundary conditions at infinity we obtain 
\be \label{sph}
    \phi_b=\sqrt\frac{2}{\l}\cdot\frac{m}{\ch{mx}}\;.
\ee
This solution is shown in Fig.~\ref{fig:sph_th}. 
Its characteristic size is $\Delta x\sim m^{-1}$ and its 
energy 
\be \label{Esph} 
    E_b=\int\diff x \left( \frac{\phi_b^{\prime \; 2}}{2} + \frac{m^2\phi_b^2}{2} -\frac{\l\phi_b^4}{4} \right) = \frac{4m^3}{3\l} \;,
\ee
sets the height of the barrier between the vacua. At $E_b\gg T$ the contribution of the critical bubble into the partition function is Boltzmann suppressed,
\be
Z_b\propto \e^{-E_b/T}=\e^{-4/(3\TT)}\;,
\ee
where in the last expression we have used the dimensionless temperature (\ref{NewTemp}). We see that this contribution is exponentially small as long as the classical theory is weakly coupled, see the condition (\ref{boundT}). In this regime $Z_b$ can be reliably evaluated in the saddle-point approximation.

Importantly, the critical bubble profile and its energy are computed using the bare parameters in the potential rather than the parameters renormalized by the thermal effects.
In other words, the fluctuations of the decaying field $\phi$ do not affect the shape or energy of the critical bubble; 
instead, their effect is captured by the fluctuation determinant~\cite{Alford:1993zf, Alford:1993ph}.
The situation is, in general, different in the presence of other fields which can significantly modify the bubble profile and its effective energy~\cite{Gleiser:1993hf, Bodeker:1993kj, Gould:2021ccf}. We do not consider this possibility in the present work, postponing it to future studies.

The critical bubble is unstable. The spectrum of its perturbations contains a unique growing mode $\vf_-\propto \e^{\omega_- t}$. Its growth rate is computed in Appendix~\ref{app:det} and reads,
\be
\label{omminus}
\omega_-=\sqrt{3} m\;.
\ee

Next we perform integration over the fluctuations around the critical bubble in the partition function (\ref{Zgen}). This completes the calculation of the critical bubble
contribution into the free energy and, combined with eq.~(\ref{omminus}), determines the prefactor in the Euclidean decay rate (\ref{G}). We relegate the details of this calculation to Appendix~\ref{app:det} and quote here the final expression:
\be
\label{G_E2}
A_E=\frac{6m^2}{\pi}\, \sqrt{\frac{E_b}{2\pi T}} \;.
\ee 
In what follows, we compare this prediction with the results of real-time numerical simulations.

\section{Classical lattice system at finite temperature}
\label{sec:num}

\subsection{Units and Hamiltonian}
\label{ssec:units}

In numerical calculations, we use the classical version of the theory (\ref{S}) discretized on a spatial lattice.
A classical thermal field theory in the continuum exhibits UV-divergences due to excitation of arbitrarily short modes (the well-known UV catastrophe). In $(1+1)$ dimensions these divergences are mild and mostly affect the total energy of the system which grows linearly with the cutoff.
The lattice spacing $a$ regularizes this growth and makes all thermodynamic averages finite.
The physical observables of interest must be independent of $a$ when the latter is much smaller than other scales in the theory. Then their values coincide with the classical limit of the corresponding observables in the continuum theory~(\ref{S}).

It is convenient to introduce the rescaled variables 
\be \label{units}
\hat{x} = mx \;, \qquad \hat{t} = mt \;, \qquad\hat\phi = (\sqrt{\l}/m)\phi \;.
\ee
This brings the action (\ref{S}) to the form 
$S=(m^2/\l) \hat{S}$ where $\hat{S}$ does not depend on the mass and coupling.
Consequently, the only parameter of the theory is the rescaled temperature $\TT$ defined in \cref{NewTemp}.
In particular, the rescaled critical bubble energy becomes a pure number, $\hat{E}_b=4/3$.
In what follows, we omit hats on the variables $\hat{x}$, $\hat{t}$ and $\hat{\phi}$, which should not lead to confusion.

We take a lattice of size $L$, with $N$ sites and periodic boundary conditions.
We label the lattice sites by $x_i\equiv a \, i$, $i=0, \dots ,N-1$, and $x_N\equiv x_0$.
The physical degrees of freedom are $\phi_i$ and their canonically conjugate momenta $\pi_i$, which we normalize by imposing the Poisson bracket $\{\phi_i,\pi_j\}=\delta_{ij}/a$.
We discretize the spatial derivative in (\ref{S}) by the second-order finite-difference operator.
Thus, the Hamiltonian of the system takes the form 
\be \label{E_d}
H =a\sum_{i=0}^{N-1}\left[ \frac{\pi_i^2}{2}-\frac{1}{2}\phi_i (\Delta \phi)_i+ \frac{\phi_i^2}{2} -\frac{\phi_i^4}{4}\right] \;,
\ee
where $(\Delta\phi)_i=a^{-2}(\phi_{i+1}-2\phi_i+\phi_{i-1})$.
It gives rise to the Hamiltonian equations of motion,
\be
\label{Hameqs}
\dot\phi_i=\pi_i\;,\qquad \dot \pi_i=(\Delta\phi)_i-\phi_i+\phi_i^3\;,
\ee
which exactly conserve the energy if the time variable is continuous. 

We numerically integrate eqs.~(\ref{Hameqs}) using the 4th order operator-splitting pseudo-spectral method (see Appendix~\ref{app:num} for details). A finite time step breaks the time-translational invariance and the exact energy conservation. We ensure that the resulting change in the value of $H$ is negligible throughout the simulations. 

Most of our simulations are performed on a grid with $L=100$, $N=8192=2^{13}$ which corresponds to a lattice spacing $a = L/N = 1.22\cdot 10^{-2}$. The time step is chosen as $h=10^{-2}$. We verified the convergence of the numerical results by varying the simulation parameters in the ranges $50\leq L\leq 400$, $5\times 10^{-3}\leq a \leq 4\times 10^{-2}$ and $0.4\leq h/a \leq 0.8$. We refer the reader to Appendix~\ref{app:num} for details. We also 
cross-checked the simulations with an independent code~\cite{Pirvu:2023plk}, which uses a 
10th order Gauss--Legendre pseudo-spectral scheme and the lattice parameters $80\leq L\leq 100$, $a=4\times 10^{-2}$, $h=0.17a$.

In what follows, we discuss the properties of the theory (\ref{E_d}) in thermal bath which are essential for the simulations of the false vacuum decay. An extensive study of the classical lattice $\l\phi^4$-theory with the positive sign of the self-coupling in $(1+1)$ and $(3+1)$ dimensions can be found in~\cite{Boyanovsky:2003tc,Destri:2004ck}.

\subsection{Equilibrium properties}
\label{ssec:equil}

\begin{figure}[t]
    \centering
    \includegraphics[width=0.6\linewidth]{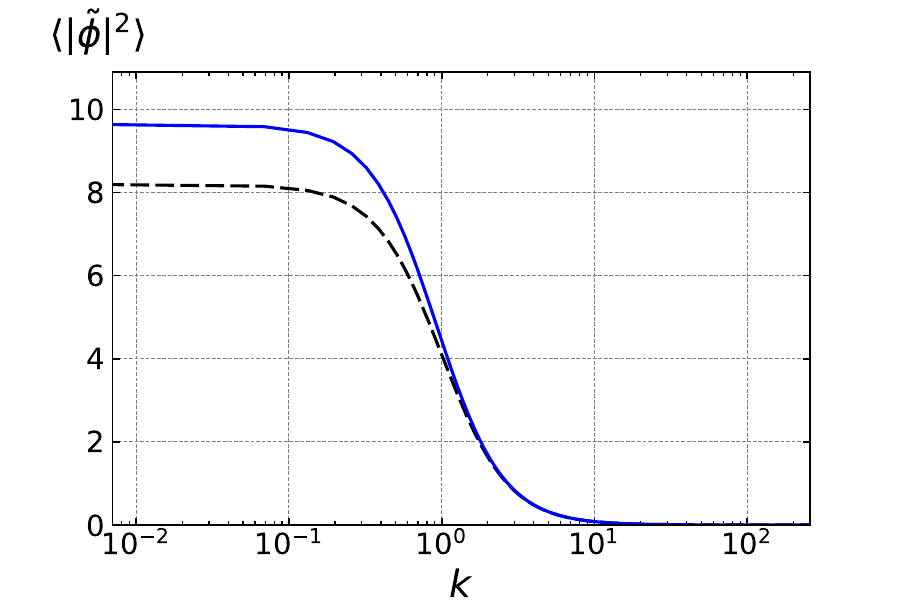}
    \caption{The tree-level (free) thermal spectrum (dashed black) and the 1-loop corrected thermal spectrum (solid blue). We take $L=100$, $N=8192$, $\TT=0.1$, which are the typical values in our simulations. The thermal correction to the spectrum is $15\%$ and must be taken into account in the initial state preparation.}
	\label{fig:ps}
\end{figure}

Since the bubble nucleation rate is exponentially sensitive to the temperature, it is crucial to set up the initial conditions for $\phi_i$ and $\pi_i$ corresponding to thermal equilibrium around the metastable state with accurately controlled temperature $\TT$. This is done by setting the appropriate Rayleigh--Jeans power spectrum for the field Fourier modes. 
We write
\be \label{IFT}
\phi_i = \frac{1}{\sqrt{N}}\sum_{j=0}^{N-1}\e^{ik_jx_i}\tilde{\phi}_j \;, \qquad k_j\equiv\frac{2\pi j}{aN}=\frac{2\pi j}{L} \;,
\ee
and similarly for $\pi_i$. The condition that the fields in coordinate space are real imposes the relations:
\be
\label{real}
\tilde\phi_j^* = \tilde\phi_{N-j}\;,\qquad \tilde\pi_j^* = \tilde\pi_{N-j}\;.
\ee
The complex amplitudes $\tilde\phi_j$, $\tilde\pi_j$ with $j=0, \dots ,[N/2]$ are then randomly 
drawn from independent Gaussian distributions with the variances\footnote{An extra care is needed for the modes with $j=0$ and $j=N/2$ lying at the boundaries of the allowed range of wavenumbers. Since they are real, their variances are reduced by half.}
\be \label{RJ}
\ev{\left|\tilde{\phi}_j\right|^2} = \frac{\TT}{a\O_j^2} \;, \quad \ev{\left|\tilde{\pi}_j\right|^2} = \frac{\TT}{a}\;.
\ee
Here, $\O_j$ are the lattice mode frequencies given by
\be \label{Oj}
\O_j^2 = \frac{2}{a^2}(1-\cos a k_j) + 1  -\frac{3\TT}{2} \;.
\ee
Note the last term in this expression, which represents the 1-loop thermal correction to the mass, see eq.~(\ref{mth}). It captures the leading effect of the self-interactions on the equilibrium spectrum. For the typical temperature in our simulations, $\TT\sim 0.1$, the correction is $\sim 15\%$ and cannot be neglected. As shown in Fig.~\ref{fig:ps}, discarding it would underestimate the power in the modes with $k_j\lesssim 1$. These modes have wavelengths comparable to the size of the bubble and one expects them to play the dominant role in the bubble nucleation. This expectation is indeed supported by the numerical data, as will be discussed in Sec.~\ref{ssec:Esph}. 

Further corrections to the spectrum are suppressed by higher powers of $\TT$ and for $\TT\sim 0.1$ are expected to be of order $1\%$, which is below the statistical uncertainty. We perform several tests to verify that this is indeed the case. We use the model with the flipped sign of the self-coupling, in which the state $\phi=0$ is stable. As a result, its evolution can be monitored for extended periods of time. First, we observe that the power spectrum set up according to eqs.~(\ref{RJ}), (\ref{Oj}) (with the flipped sign of the last term in (\ref{Oj})) remains stable over time, within the statistical fluctuations. In contrast, omitting the thermal mass correction in (\ref{Oj}) leads to pronounced oscillations of the spectrum.

\begin{figure}[t]
    \centering
    \includegraphics[width=0.6\linewidth]{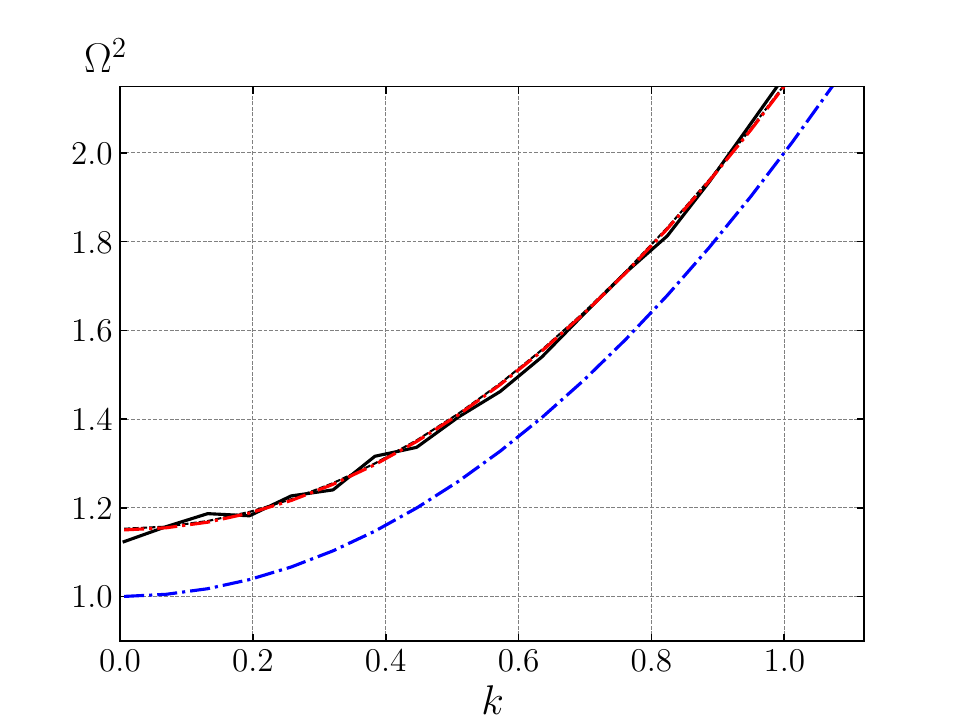}
    \caption{Measurement of the thermal mass using 100 realizations of the Langevin dynamics with $\heta=1$ and $\TT=0.1$. Mode frequencies are estimated using eq.~(\ref{app:mth}). For each realization, the spectra are sampled at 100 time moments evenly distributed in the interval $0<t<100$. The simulation data are shown by the solid black curve. The black dashed line shows their best fit using eq.~(\ref{disp}). It is in excellent agreement with the one-loop prediction (red dot-dashed curve). The tree-level dispersion relation (blue dot-dashed curve) does not describe the data.
    }
	\label{fig:app:meff}
\end{figure}

Second, we model the coupling of the system to an external heat bath by replacing the Hamiltonian equation with the Langevin dynamics (to be discussed in detail in Sec.~\ref{sec:simL}). We choose a strong dissipation $\heta=1$ which quickly brings the system to thermal equilibrium, irrespective of the initial conditions, and measure the resulting thermal mass of the long modes. To this end, we evaluate the ratio of the $\pi$ and $\phi$ power spectra averaged over an ensemble of 100 realizations with $\TT=0.1$,
\be \label{app:mth}
\Omega^2_{j}=\frac{\ev{ \left|\tilde{\pi}_j \right|^2 }}{\ev{ \left|\tilde{\phi}_j \right|^2 }} \;, \qquad k_j \ll \frac{1}{a} \;.
\ee
This ratio provides an estimator of the mode frequencies. Plotting it against $k_j$ we obtain the effective dispersion relation of long modes in the thermal environment. The result is shown in Fig.~\ref{fig:app:meff} by the black solid curve. Clearly, it significantly deviates from the tree-level dispersion relation $\Omega_j^2=k_j^2+1$ (blue dot-dashed curve). On the other hand, it is well fitted by the Ansatz
\be
\label{disp}
\Omega_j^2=k^2+\hat{m}_{\rm th}^2\;,
\ee
with $\hat{m}_{\rm th}^2=1.153$. The latter value is less than $1\%$ off the 1-loop thermal mass $\hat{m}_{\rm th}^2=1.150$ for the temperature $\TT=0.1$.

Additional corrections to the thermal spectrum arise due the finite box size and lattice spacing. However, as discussed in Appendix~\ref{app:num}, these are completely negligible.

Before closing this subsection, let us introduce the notion of an 
`effective temperature' of Fourier modes within a certain range of wavenumbers
$k_{\rm min}<k_j<k_{\rm max}$. We define it through the average amplitude of the momentum power spectrum in the range,
\be \label{Teff}
\Teff = a \ev{\left|\tilde{\pi}_j\right|^2}_{k_{\rm min}<k_j<k_{\rm max}} \;.
\ee
This observable is useful when the system deviates from equilibrium, and the difference between $\Teff$ and $\TT$ is a measure of such deviation.

\subsection{Thermalization of long modes}
\label{ssec:therm}

Long modes with $k\sim 1$ comprising the critical bubble propagate in the background of short modes. Since in equilibrium each mode carries approximate energy $\TT$, and number of short modes is much larger, they dominate the total energy of the system. Thus, they provide a heat reservoir for the long modes. The efficiency of this reservoir in maintaining thermal equilibrium is limited by the speed of energy transfer between different modes characterized by the thermalization time $t_{th}$. 
In this subsection, we estimate the thermalization time in the model at hand both analytically and numerically. 

For the analytic estimate, we consider the field as a collection of quantum particles and write the Boltzmann equation for particle phase-space density $f_p$. The classical dynamics corresponds to the leading order in this equation, 
see e.g.~\cite{Mueller:2002gd}.
It is convenient to revert back the physical units for a moment and restore explicitly the field mass and coupling in the equations. Due to the features of the $(1+1)$-dimensional kinematics, the ${2\leftrightarrow 2}$ scattering preserves the energy distribution, so the leading 
processes resulting in the energy exchange between the particles are the $2\leftrightarrow 4$ and $3\leftrightarrow 3$ scatterings. They give comparable contributions into the collision integral. For concreteness, let us focus on the former. Denoting the momenta of incoming particles by $p_1$, $p_2$ and the momenta of the outgoing particles by $p_3,p_4,p_5,p_6$ we have,
\be
    \begin{split}
        \frac{\d f_{p_1}}{\d t} \simeq & \frac{1}{2\o_{p_1}}\int\frac{\diff \vec{p}_2\diff\vec{p}_3\diff \vec{p}_4\diff \vec{p}_5\diff \vec{p}_6}{(2\pi)^5 2\o_{p_2}2\o_{p_3}2\o_{p_4}2\o_{p_5}2\o_{p_6}}  (2\pi)^2\delta^{(2)}(p_1+p_2-p_3-p_4-p_5-p_6) \left|\mathcal{A}_{2\to 4}\right|^2 \\
        & \times \Bigl[ - f_{p_1}f_{p_2}(1+f_{p_3})(1+f_{p_4})(1+f_{p_5})(1+f_{p_6}) + (1+f_{p_1})(1+f_{p_2})f_{p_3}f_{p_4}f_{p_5}f_{p_6}  \Bigr],
    \end{split}
\ee
where $\omega_{p_1}$, etc., are the particles' energies. We assume that the distribution is slightly perturbed around its equilibrium value, $f_p=T/\omega_p+\delta f_p$. Let us further
assume for simplicity that all particles have comparable momenta of order $p$. Then, the scattering amplitude is $\mathcal{A}_{2\to 4}\sim \l^2/\o_p^2$,
and the dominant contribution from the Bose-enhancement factor is $f_p^4\delta f_p \sim (T/\o_p)^4 \delta f_p$. This yields, 
\be \label{cross-section}
    \frac{1}{\delta f_p}\frac{\d \delta f_p}{\d t}\sim \frac{1}{(2\pi)^3}\frac{\l^4T^4}{\o_p^{11}}\;,
\ee
whence we read off the thermalization time  
\be \label{t_therm2}
    t_{th} \sim \frac{(2\pi)^3}{m} \left(\frac{m^3}{\l T} \right)^4\left(\frac{\o_p }{m} \right)^{11} \, .
\ee
Note the steep increase of $t_{th}$ with the particle's energy.
The modes relevant for the decay have energies $\o_p\sim m$.
Their thermalization predominantly takes place through the interaction with modes of comparable energy.
Substituting $\o_p\sim m$ in \cref{t_therm2} and switching back to the dimensionless units, we obtain
\be \label{t_therm3}
t_{th}\sim\frac{(2\pi)^3}{\TT^4} \;.
\ee 
At weak coupling, $\TT\ll 1$, the thermalization time is much longer than the dynamical time of the theory set by the inverse field mass ($m=1$ in the dimensionless units).

\begin{figure}[t]
    \begin{minipage}[h]{0.49\linewidth}
        \centering
        \includegraphics[width=1.0\linewidth]{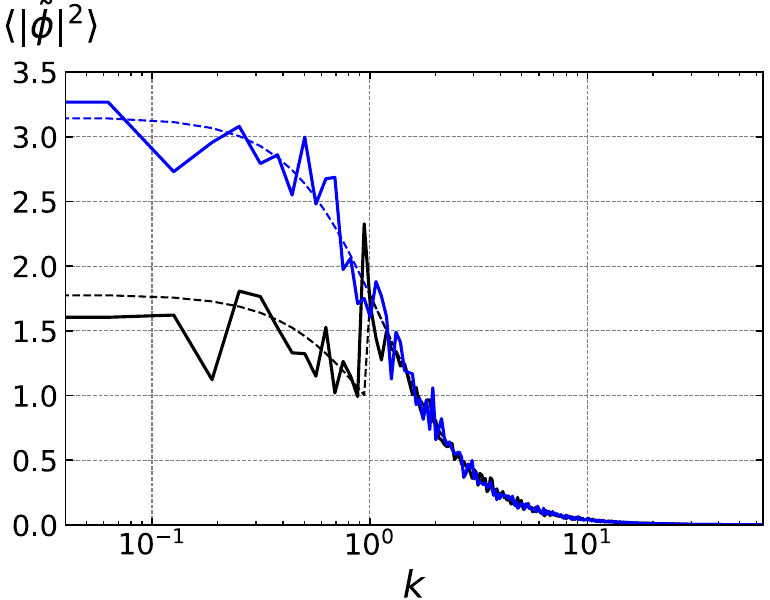} \\ (a)
	\end{minipage}
	\begin{minipage}[h]{0.49\linewidth}
        \centering
        \includegraphics[width=1.0\linewidth]{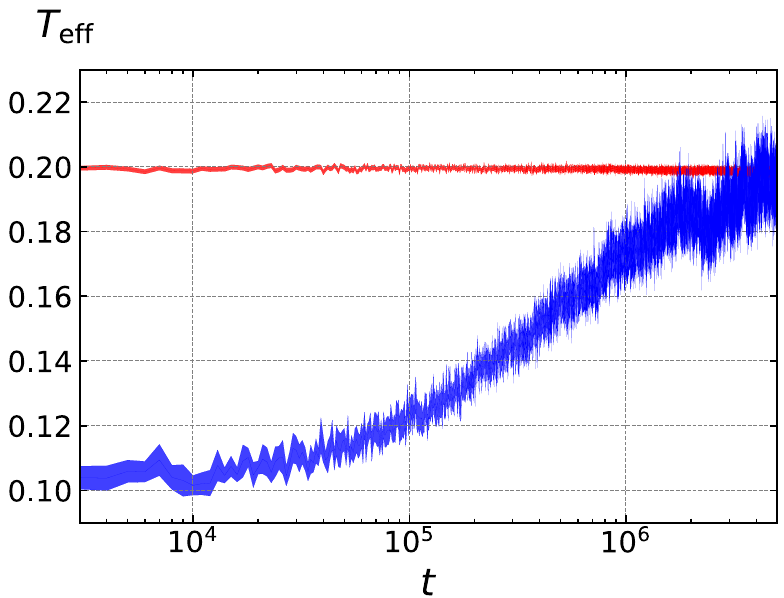} \\ (b)
	\end{minipage}
    \caption{Thermalization of long modes in the classical lattice theory with positive quartic coupling.   
    \textbf{(a)} The initial spectrum with the power of long modes artificially reduced by half (black) and the final thermal spectrum (blue) at $t=5\cdot 10^6$. \textbf{(b)} The effective temperature (\ref{Teff}) of the long modes with $k<1$ (blue) and of the modes with $k>1$ (red). The width of the lines represents statistical uncertainty.
    For both plots we take $L=100$, $N=2048$ and average over 100 realizations of the initial spectrum.}
	\label{fig:therm}
\end{figure}

To check the estimate (\ref{t_therm3}), we make the following numerical experiment.
We again use the model with positive quartic coupling to switch off the decays. 
We prepare an ensemble of initial states according to the equilibrium distribution (\ref{RJ}), (\ref{Oj}) with $\TT=0.2$ 
(in \cref{Oj} we flip the sign of the last term).
Then, the power in the long modes with $k<1$ is reduced artificially by a factor of 2, so that their effective temperature becomes $\Teff = a\langle \left|\tilde{\pi}_j\right|^2\rangle_{k<1}=0.1$.
The modified spectrum is shown in Fig.~\ref{fig:therm}(a) by the black curve.
We evolve the resulting ensemble and observe how the energy transfer from the modes with $k>1$ restores the power in the long modes.
The result of the experiment is shown in Fig.~\ref{fig:therm}(b).
We see that $\Teff$ is driven to the equilibrium value set by the short modes
on the time scale 
$t\sim (\text{a few})\times 10^6$, in agreement with \cref{t_therm3}.\footnote{In this 
estimate we use $T_{\rm eff}$, rather than $\TT$, since the former sets the initial temperature of the long modes.} 

These results are consistent with the previous numerical studies of thermalization in the $\l\phi^4$-theory which also reveal
very long thermalization times~\cite{Boyanovsky:2003tc, Destri:2004ck}.
Note that slow thermalization is not a peculiarity of the $(1+1)$-dimensional theory, but is a feature of any single-field theory at weak coupling.
For example, the $\l\phi^4$-theory in $(3+1)$-dimensions exhibits similar behavior~\cite{Destri:2004ck}.

We conclude that the model (\ref{S}) does not admit an efficient energy exchange between the heat reservoir and the subset of modes comprising the bubble. As we will see shortly, this has implications for the dynamics and rate of bubble nucleation.

\section{Thermal transitions in the Hamiltonian system}
\label{sec:sim}

\subsection{Critical bubble profile}
\label{ssec:sph}

The standard theory of false vacuum decay predicts the shape of the critical bubble. 
We would like to reconstruct it from our simulations and compare with \cref{sph}.
We use two independent reconstruction methods.
In each method we prepare an ensemble of states according to \cref{RJ}.
The states are then evolved until a decay is detected (or until the simulation times out).
The decay event is identified when the maximal absolute value of the field in the simulation exceeds a threshold, $\left|\phi\right| > 10$. 
We record the evolution of the field preceding the decay starting from the time before the bubble nucleation. 
Next, we synchronize all field histories with respect to the decay event, and produce the averaged field evolution across the barrier.
Finally, we pinpoint the time slice corresponding to the critical bubble in the average history. 

In Method 1 the above steps are performed by studying the evolution of long modes with $k\lesssim 1$.
Their potential energy $U_{long}(t)$ (which includes the gradient energy) switches from positive to negative values shortly after the system passes over the barrier.
Once $U_{long}(t)$ becomes negative, i.e. after the critical bubble formation, its behavior in different simulations becomes very similar since it is determined by the bubble growth, which is universal. 
We shift the recorded simulations in time, relative to each other, to achieve the maximal correlation between the negative asymptotics of all functions $U_{long}(t)$.
To centralize the decay events in space, we correlate the field profile in each simulation against a mock Gaussian profile, making the reflection $\phi\mapsto-\phi$ when necessary.
Finally, we average all simulations and obtain the averaged history $\phi_{av}(t,x)$.

\begin{figure}[t]
    \begin{minipage}[h]{0.49\linewidth}
        \centering
        \includegraphics[width=1.0\linewidth]{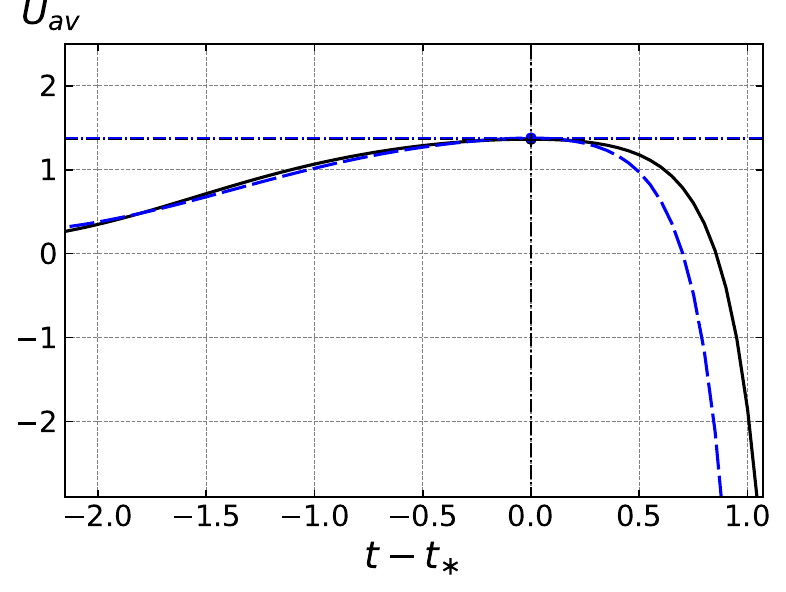} \\ (a)
	\end{minipage}
	\begin{minipage}[h]{0.49\linewidth}
        \centering
        \includegraphics[width=1.0\linewidth]{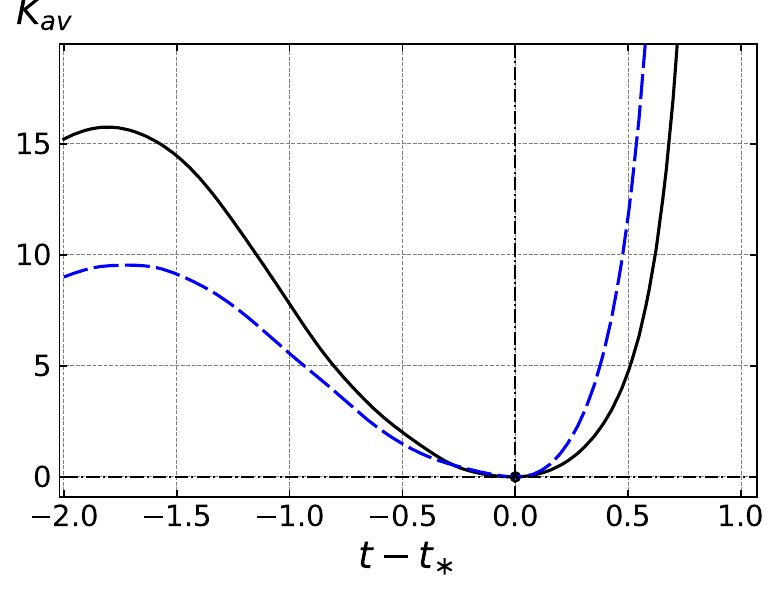} \\ (b)
	\end{minipage}
    \caption{\textbf{(a)} Evolution of the potential energy $U_{av}$ of the averaged field in Method 1 at $\TT=0.1$ (solid black), $\TT=0.2$ (dashed blue). The maximum of $U_{av}$ determines the moment of bubble nucleation $t_*$. 
    \textbf{(b)} Evolution of the kinetic energy $K_{av}$ of the averaged field in Method 2 for the same values of temperature. The moment of bubble nucleation $t_*$ is identified with the minimum of $K_{av}$. In both methods the averaging is performed over 250 simulations.
    }
	\label{fig:Uav(t)}
\end{figure}

\begin{figure}[t]
	\begin{minipage}[h]{0.49\linewidth}
        \centering
        \includegraphics[width=1.0\linewidth]{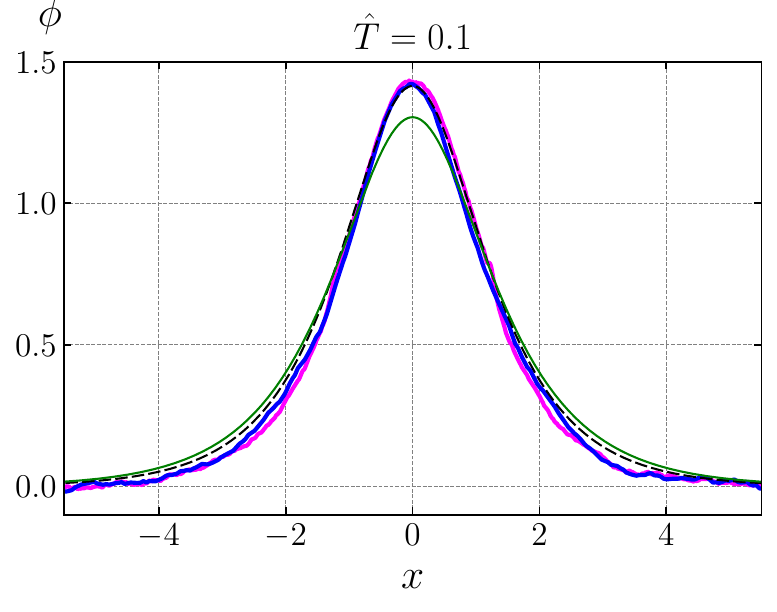}
	\end{minipage}
	\begin{minipage}[h]{0.49\linewidth}
        \centering
        \includegraphics[width=1.0\linewidth]{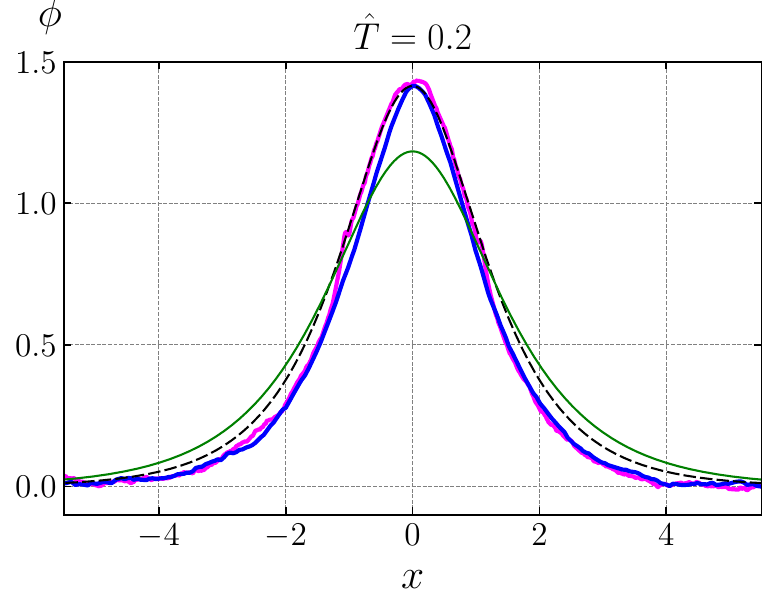}
	\end{minipage}
    \caption{Critical bubble profile reconstructed from simulations using Method 1 (solid magenta) and Method 2 (solid blue). 
    The dashed black line is the Euclidean prediction (\ref{sph}). The solid green line corresponds to \cref{sph} with the bare field mass replaced by the thermal mass (\ref{mth}). }
	\label{fig:Sph(t)}
\end{figure}

Fig.~\ref{fig:Uav(t)}(a) shows the potential energy of the averaged field.
We see that it reaches a local maximum $U_{\rm max}$ before plunging to negative values.
We identify this moment with the nucleation of the critical bubble,
and the averaged field profile at this moment with the critical bubble profile. Note that 
$U_{\rm max}$ provides an estimate of the critical bubble energy. We measure $U_{\rm max} \approx 4/3=\hat{E}_b$ within the accuracy of the numerical routine, in agreement with the theoretical expectation.\footnote{In Sec.~\ref{sec:comparison} we will measure the critical bubble energy with a higher precision using a different observable.}
The reconstructed bubble profile is shown in Fig.~\ref{fig:Sph(t)} by the solid magenta line. 
We observe that it agrees well with the 
profile (\ref{sph}) predicted by the Euclidean theory. A mild discrepancy at the level $\sim 10^{-2}$ observed at the tails can be due to a bias introduced by an imperfect filtering out of the short-wavelength fluctuations (noise).

To make sure that the details of the synchronization procedure do not affect the result, we validate it with an alternative  
bubble reconstruction routine developed in~\cite{Pirvu:2023plk} (Method~2).
The reader is referred to that work for its detailed description.
It accounts for the fact that, in general, the center-of-mass of the nucleated bubble moves in the rest frame of the thermal bath.
Hence, a de-boosting procedure is first applied to place all bubbles at rest.
Next, the synchronization of the decay events is achieved by studying the motion of the bubble walls after the nucleation.\footnote{To study the field dynamics after nucleation, a higher-order coupling term is added to the theory (\ref{S}) to place the true vacuum at a finite field value. A small value of the new coupling ensures that the critical bubble is unchanged.}
As they accelerate, the bubble walls gain large momenta and get Lorentz-contracted, thus becoming insensitive to thermal fluctuations.
Due to this, the bubble expansion at late times is fully determined by hyperbolic bubble wall trajectories.
The center of the hyperbola is then identified with the nucleation site, with respect to which we synchronize the recorded simulations.
Finally, the averaged history $\phi_{av}(t,x)$ is produced.

Fig.~\ref{fig:Uav(t)}(b) shows the kinetic energy $K_{av}=\int dx \,\dot\phi_{av}^2/2$ 
of the averaged field produced with the second method.
We identify its local minimum with the moment of the bubble nucleation.\footnote{
For presentation purposes, $K_{av}$ is shifted by a constant to set its value at the minimum to zero. The local maximum of $K_{av}$ observed at an earlier time also has a physical meaning: it reveals oscillation of the kinetic energy in the oscillonic precursor that leads to the critical bubble, see Sec.~\ref{sec:osc} for more details. }
The corresponding field profile is shown in Fig.~\ref{fig:Sph(t)} by a solid blue line. 
We see a remarkable agreement with the profile obtained by the first method. Since the two reconstruction procedures are independent and use different numerical tools, we conclude that the measured critical bubble profile is robust. 

The theoretical bubble profile (\ref{sph}) is found using the bare field equation (\ref{spheq1}). On the other hand, as discussed in Secs.~\ref{ssec:effpot},~\ref{ssec:equil}, the field receives a sizable thermal correction to the mass. One may wonder if this should be included when determining the bubble profile. To test this possibility, we compute the bubble profile from \cref{spheq1} where we replace the bare mass $m$ with the thermal mass $m_{\rm th}$. The result is shown in green in Fig.~\ref{fig:Sph(t)}, and is clearly inconsistent with the data. This confirms the assertion~\cite{Alford:1993ph, Alford:1993zf, Gleiser:1993hf} that in single-field theories the critical bubble shape is unaffected by the classical thermal fluctuations.\footnote{By contrast, \textit{quantum} thermal corrections (hard thermal loops) 
can significantly alter the tunneling potential and the shape of the critical bubble. They are absent in our $(1+1)$-dimensional model but can be relevant in $(3+1)$ dimensions, where they can even be responsible for the appearance or disappearance of the metastable vacuum. The contribution of 
hard thermal loops must be included into the `bare' action of the classical theory \cite{Parwani:1991gq,Bodeker:1993kj,Aarts:1997kp,Gould:2021ccf}. }

\subsection{Classical Zeno effect}
\label{ssec:Esph}

We proceed to measuring the decay rate. We run an ensemble of simulations with the initial conditions (\ref{RJ}) at a fixed temperature $\TT$ and count the number of surviving simulations (i.e. simulations that have not yet decayed) as function of time. The  survival probability $P_{surv}(t)$ is then defined as the ratio of this number to the total initial number of simulations in the ensemble. For an exponential decay, the survival probability is expected to have the form,
\be \label{Psurv}
\ln P_{surv}(t)=\const-\G L\cdot t \;,
\ee
where the constant term accounts for possible transients at the beginning of the simulations. In other words, logarithm of $P_{surv}(t)$ is expected to be a straight line, with the slope proportional to the decay rate. 

We run simulations and measure the curves $\ln P_{surv}(t)$ in a range of temperatures ${0.09\leqslant \TT \leqslant 0.13}$.
The lower limit here is determined by the requirement to have a sufficiently large number of decays within a reasonable runtime: for temperatures below $0.09$ the decay rate is too low to admit enough decay statistics. The upper limit is chosen so that 
$\Es/\TT>10$ and the higher-loop corrections in the Euclidean calculation of the decay rate are suppressed. 

\begin{figure}[t]
    \centering
    \includegraphics[width=0.6\linewidth]{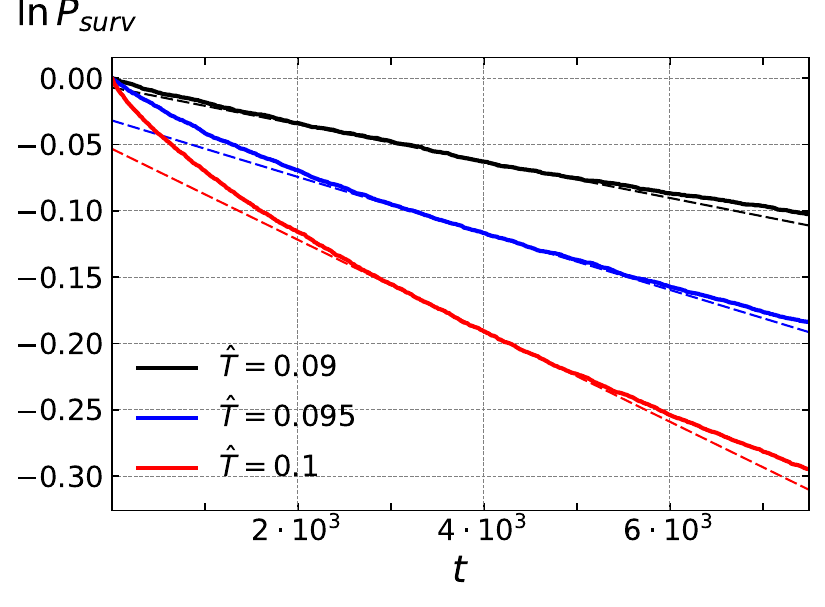}
    \caption{Survival probability measured in ensembles of numerical simulations with three different temperatures. 
     Straight dashed lines are tangent to the probability curves at $t=4\cdot 10^3$ and serve to illustrate the time dependence of their slopes. }
	\label{fig:Psurv}
\end{figure}

The measured survival probability is shown in 
Fig.~\ref{fig:Psurv} for several values of $\TT$.
We immediately see that the slope of $\ln P_{surv}(t)$ is not constant, in contradiction with the expectation (\ref{Psurv}). The flattening of the curves implies the decrease of the decay rate with time. This decrease sets early on --- it becomes visible already when a small fraction 
($\approx 5\%$) of the ensemble decays --- and persists into the times when the decayed fraction becomes of order one. The latter property excludes its interpretation as a transient phenomenon associated with the start of the simulations.

To understand what is going on, we recall that the thermalization time in the system is very long, $t_{th}\sim 10^6$ (see Sec.~\ref{ssec:therm}). It exceeds the decay time $t_{dec}\sim (\G L)^{-1}$
which in our simulations is between $10^3$ and $10^5$, depending on the temperature. This means that the initial distribution of energy between different Fourier modes in a given simulation is essentially preserved throughout the whole evolution until its decay. Since the critical bubble has unit size in the dimensionless units ($m^{-1}$ in the physical units), it is natural to conjecture that its formation probability is mainly determined by the power in the long modes with wavenumbers $k\lesssim 1$. This power is conveniently characterized by the effective temperature $\Teff^{\rm long}$ defined through \cref{Teff} with $k_1=0$ and $k_2\sim 1$. Simulations which, due to random fluctuations in the initial data, have higher $\Teff^{\rm long}$ will decay faster, whereas the simulations with lower $\Teff^{\rm long}$ will survive longer. This biases the statistical properties of the ensemble as the time proceeds: since the realizations with high $\Teff^{\rm long}$ are washed out by decays, the average effective temperature $\Teff^{\rm long}$ in the ensemble decreases. This, in turn, reduces the decay rate. 

\begin{figure}[t]
    \begin{minipage}[h]{0.49\linewidth}
        \centering
        \includegraphics[width=1.02\linewidth]{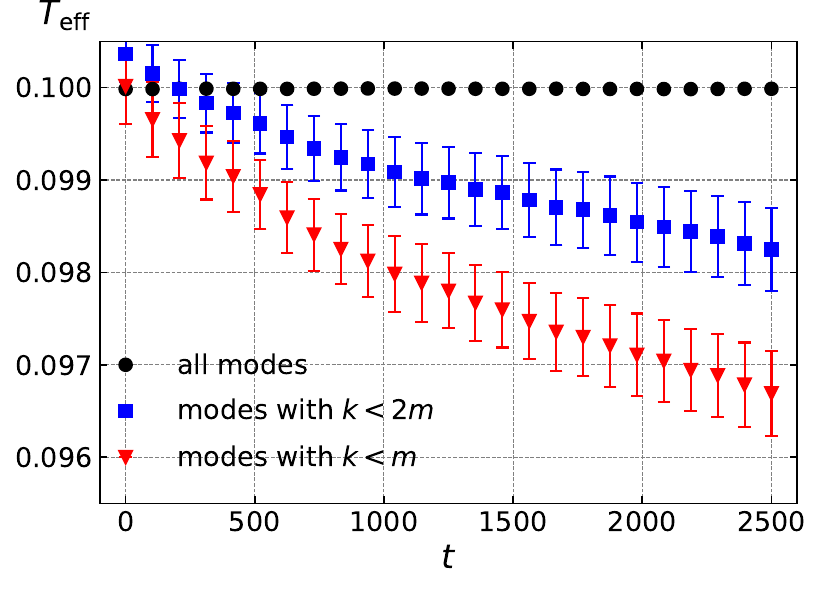} \\ (a)
	\end{minipage}
	\begin{minipage}[h]{0.49\linewidth}
        \centering
        \includegraphics[width=1.02\linewidth]{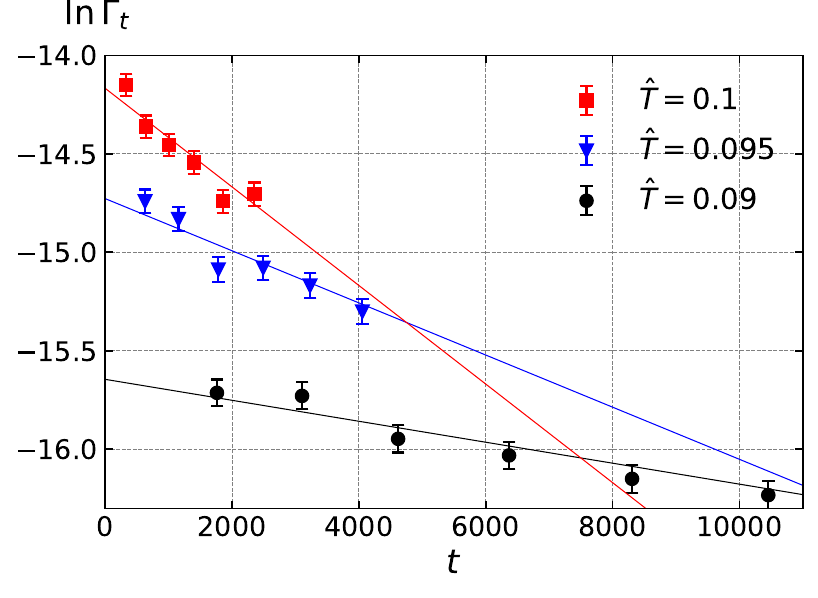} \\ (b)
	\end{minipage}
    \caption{ \textbf{(a)} Effective temperature of long modes, averaged over the surviving configurations at time $t$. The temperature computed using all modes is also shown and is time-independent. 
    \textbf{(b)}~Time dependence of the decay rate for several values of temperature. Solid lines show the linear fit. In both plots error bars represent the statistical uncertainty.
    }
	\label{fig:Teff}
\end{figure}

We test this picture by measuring $\Teff^{\rm long}$ in the ensemble of surviving configurations as function of time. The result is shown in Fig.~\ref{fig:Teff}(a) and demonstrates a clear decrease in the long-mode effective temperature by a few per cent compared to the initial value. Note that this effect becomes completely invisible if we measure the temperature averaged over all modes. We view this as a confirmation that the decay rate is sensitive to $\Teff^{\rm long}$, rather than the overall temperature of the ensemble.
Let us approximate the effective temperature as 
$\Teff^{\rm long}\simeq \TT\,(1-\alpha t)$ where $\alpha\sim \Gamma L$.
Then, the time-dependent decay rate at small $\alpha t$ can be estimated as
\be \label{G(t)}
\G_t\simeq\Gamma \,\e^{- (\alpha \Es/\TT)\,t} \;.
\ee
Since $\Es/\TT\gg 1$, the effect becomes significant even for $\G L\, t\ll 1$, in agreement with the results presented in Fig.~\ref{fig:Psurv}.

To sum up, in cases when $t_{dec}<t_{th}$ the vacuum decay happens to be non-Markovian: 
a system that is observed not to decay within a given time has a lower chance of decaying in the future. 
This resembles the Zeno effect, which allows one to freeze the evolution of a quantum system by measuring its state.
We stress, however, that in our case, the effect is purely classical.

Since the thermalization time $t_{th}$ depends on the inverse temperature only polynomially, whereas $t_{dec}$ grows exponentially, the condition $t_{dec}\gg t_{th}$ will eventually be fulfilled at sufficiently low temperatures. In this case the classical Zeno effect is expected to disappear. However, reaching this regime in practice may be challenging. This applies both to numerical simulations, like our own, as well as to laboratory experiments, especially those using $(1+1)$-dimensional systems.

\subsection{Unbiased rate}
\label{ssec:unbiased}

The classical Zeno effect described above is not captured by the Euclidean formula (\ref{G}). By construction, the latter is supposed to provide the decay rate of an unbiased canonical ensemble with fixed temperature $\TT$. Such an ensemble is present in our simulations only at the initial moment of time. Thus, for a detailed comparison with the Euclidean theory, we need to measure the decay rate at $t=0$, which corresponds to the slopes of the curves in Fig.~\ref{fig:Psurv} at the origin.

Measuring the rate at small $t$ directly would suffer from large statistical uncertainties and transient effects. Instead, we measure the rate at a sequence of points $\bar{t}_i$ and then extrapolate the result to $t\to 0$. In more detail, we split the probability curve 
$y=\ln P_{surv}(t)$ into segments located within the intervals $(t_0=0,t_1)$, $(t_1,t_2),\dots$, such that the fraction of decays in each segment is small (a few per cent) and is the same for all segments, 
$y_0-y_1=y_1-y_2=\dots$.
We remove the first segment from the analysis to avoid transients.
The remaining segments of the curve are approximated by straight lines.
Denote by $(-\G_i)$ the slope of the line fitting the $i^{th}$ segment.
We identify $\ln\G_i$ as the logarithm of the decay rate at $\bar{t}_i=(t_i+t_{i+1})/2$. This gives a set of points $(\bar{t}_i, \ln\G_i)$ which are fitted by a linear dependence, see Fig.~\ref{fig:Teff}(b). The intercept of this dependence is taken as the logarithm of the unbiased rate $\ln\G$. 
We repeat the above procedure for nine values of temperature spanning the interval $0.09\leqslant \TT\leqslant 0.13$ and
obtain the function $\G(\TT)$.

The uncertainty of this measurement is estimated as follows. The variance of the rate $\G_i$ in each time segment is given by the Poisson statistics (see Appendix~\ref{app:stat}), 
\be \label{sigma_lnG}
\sigma_{\ln\G_i} = \frac{\sigma_{\G_i}}{\G_i}=\frac{1}{\sqrt{N_i}} \;,
\ee
where $N_i$ is the number of decay events in the segment. In our case $N_i\sim 100$ yielding $\sigma_{\ln \G_i}\sim 0.1$.
The error of the extrapolated value $\s_{\ln\G}$ is then obtained from the errors $\s_{\ln\G_i}$ using the linear regression.

\subsection{Comparison with the Euclidean prediction}
\label{sec:comparison}

We fit the temperature dependence of the unbiased rate 
by the expression
\be \label{log_G(T)}
    \ln\G(\TT) =- \frac{1}{2}\ln \TT + \ln A - \frac{B}{\TT} \;,
\ee
with free parameters $A$ and $B$. The first term on the right accounts for the temperature dependence of the prefactor predicted by \cref{G_E2}.

\begin{figure}[t]
    \begin{minipage}[h]{0.49\linewidth}
        \centering
        \includegraphics[width=1.0\linewidth]{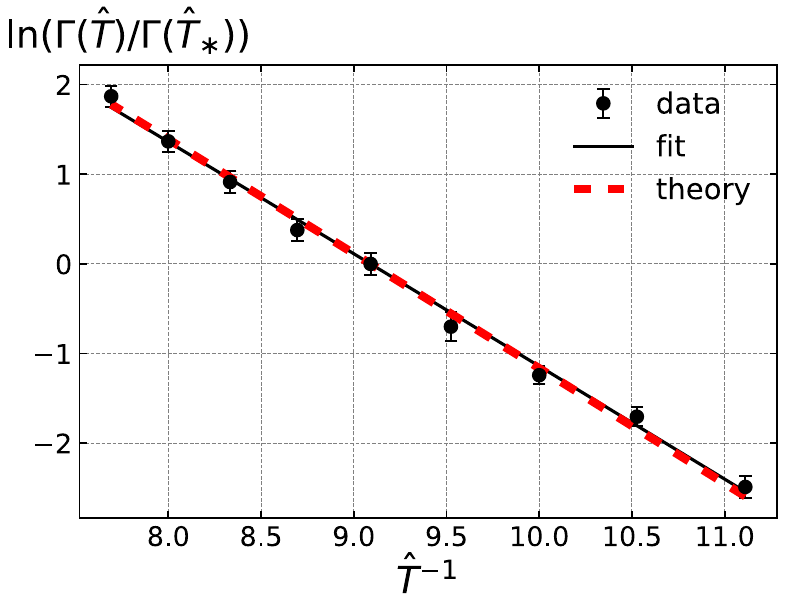} \\ (a)
	\end{minipage}
	\begin{minipage}[h]{0.49\linewidth}
        \centering
        \includegraphics[width=1.0\linewidth]{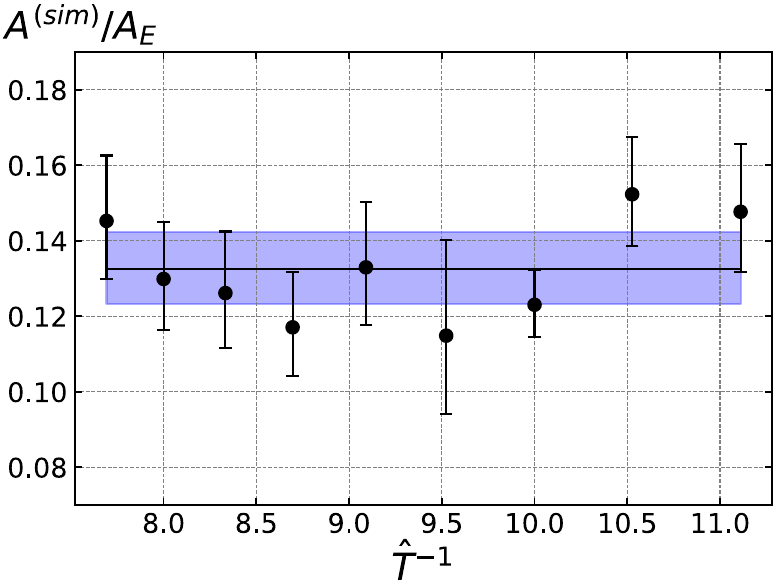} \\ (b)
	\end{minipage}
    \caption{ \textbf{(a)} Measuring the exponential suppression of the decay rate. The unbiased rate normalized at a pivot point $\TT_*=0.11$ is shown as function of the inverse temperature (black dots). The black solid line corresponds to \cref{log_G(T)} with the best-fit value of $B$. The red dashed line is the prediction of Euclidean theory.
    \textbf{(b)} Prefactor measured from the simulations normalized to the prediction of Euclidean theory at different temperatures (black dots). The black horizontal line and the blue band show the mean and the 
    uncertainty of the measurement.}
	\label{fig:E_sph}
\end{figure}

First, we measure the value $B$ associated with the leading exponential suppression. 
To this end, we eliminate the parameter $A$ by taking 
the ratio $ \ln[\Gamma (\TT) / \G(\TT_*)]$ with the pivot point in the middle of the studied temperature range, 
$\TT_* = 0.11$.
The corresponding one-parameter fit is shown in Fig.~\ref{fig:E_sph}(a). 
It yields $B/\Es = 0.98\pm 0.02$, implying that the exponential suppression is given by the critical bubble energy within $2\%$ precision. This agrees with the Euclidean theory prediction. Note that $\Es$ does not receive any thermal corrections, cf.~\cite{Alford:1993zf,Gleiser:1993hf, Alford:1993ph}.

Next, we measure the parameter $A$ entering the prefactor.
We equate $B$ to its theory value $\Es$ in \cref{log_G(T)} and extract $A$ at different values of temperature. 
The ratio of the result to the Euclidean prediction of \cref{G_E2} is shown in Fig.~\ref{fig:E_sph}(b).
It is temperature-independent within the error bars.
Thus, we can combine the data at different $\TT$ and obtain
\be \label{Aration}
    A^{(sim)}/A_E=0.13 \pm 0.01 \;.
\ee
We conservatively estimate the uncertainty as the maximal possible systematic error of the measurements, which for each individual temperature is subdominant to the statistical error (see Appendix~\ref{app:num}). 
We see a factor $\sim 8$ discrepancy in the prefactor between the real-time simulations and the Euclidean theory.
We have verified that a full two-parameter fit of the function $\G(\TT)$ by \cref{log_G(T)}, with both $A$ and $B$ allowed to vary freely, yields a best-fit value for $A^{(sim)}/A_E$ consistent with \cref{Aration}, with the error bar increased by a factor $\sim 2$. 

The observed discrepancy (\ref{Aration}) cannot be attributed to two-loop corrections within the Euclidean approach, since the latter are expected to affect the prefactor only at the $\TT\sim 10\%$ level.
It cannot be due to the classical Zeno effect either, since the latter has been mitigated by constructing the unbiased decay rate; furthermore, the whole change of $\G_t$ over the time span used in this construction is less than a factor of $2$.
Finally, as will be discussed in the next section, it cannot be due to some subtle deviations from thermality in the initial data.

We are forced to conclude that the Euclidean theory fails to describe the prefactor in the bubble nucleation rate. The likely origin of this failure has been discussed in the Introduction: it is the large hierarchy between the short dynamical time scale of the bubble and the long thermalization time, which leads to the violation of thermal equilibrium during the bubble nucleation. We stress that this deviation from equilibrium is distinct from the classical Zeno effect. While the latter pertains to the statistical properties of the ensemble, the former is dynamical and occurs in every individual realization, whenever a critical bubble forms. 

To support this interpretation and explore the dependence of the rate on the thermalization time, we now couple the theory (\ref{S}) to an {\it external} heat bath. The latter is modeled by promoting the field equation of motion to the stochastic Langevin equation.

\section{Thermal transitions with the Langevin evolution}
\label{sec:simL}

\subsection{Reduced thermalization time}

To model the effect of an external ideal thermostat, we introduce stochastic terms into the field's classical equation of motion.
Namely, we promote it to the Langevin equation 
\be \label{LangEq}
\ddot\phi + \heta\dot\phi - \phi^{\prime\prime} + \phi - \phi^3 = \xi \;,
\ee 
where $\heta\equiv\eta/m$ is the dimensionless 
dissipation coefficient and $\xi=\xi(t,x)$ is a white noise satisfying
\be \label{Noise}
\ev{\xi(t,x)} = 0 \;, \quad \ev{\xi(t,x) \xi(t^{\prime},x^{\prime})} = 
2\heta \TT\,\delta(t-t^{\prime}) \delta(x-x^{\prime}) \;,
\ee
in agreement with the fluctuation-dissipation theorem.
Every Fourier mode of the field $\phi$ is now coupled to the external heat bath, 
and its thermalization rate is set by the dissipation coefficient rather than the interaction with other modes:
\be \label{t_therm_eta}
t_{th}\sim\heta^{-1} \;.
\ee
This is true as long as $\heta$ is larger than the inverse thermalization time in the Hamiltonian system, $\heta\gtrsim 10^{-6}$.

\begin{figure}[t]
    \begin{minipage}[h]{0.49\linewidth}
        \centering
        \includegraphics[width=1.0\linewidth]{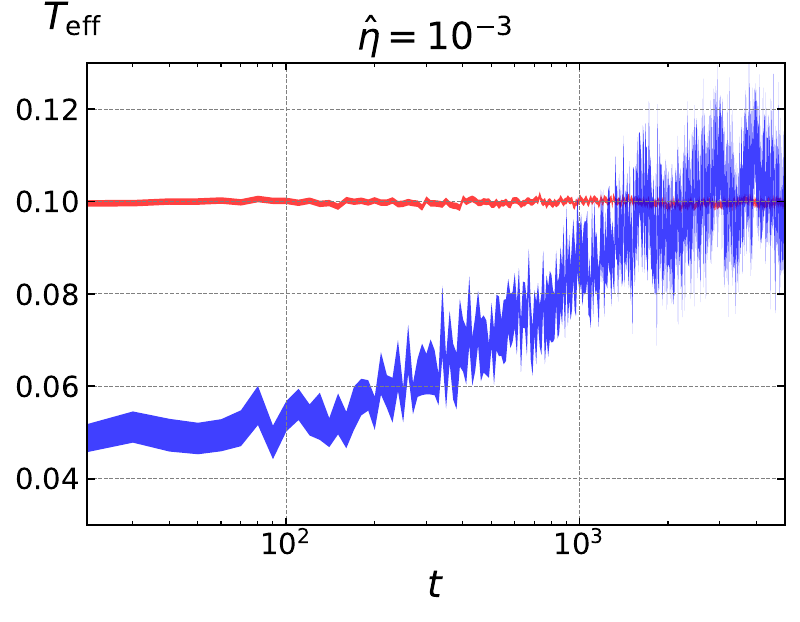}
	\end{minipage}
	\begin{minipage}[h]{0.49\linewidth}
        \centering
        \includegraphics[width=1.0\linewidth]{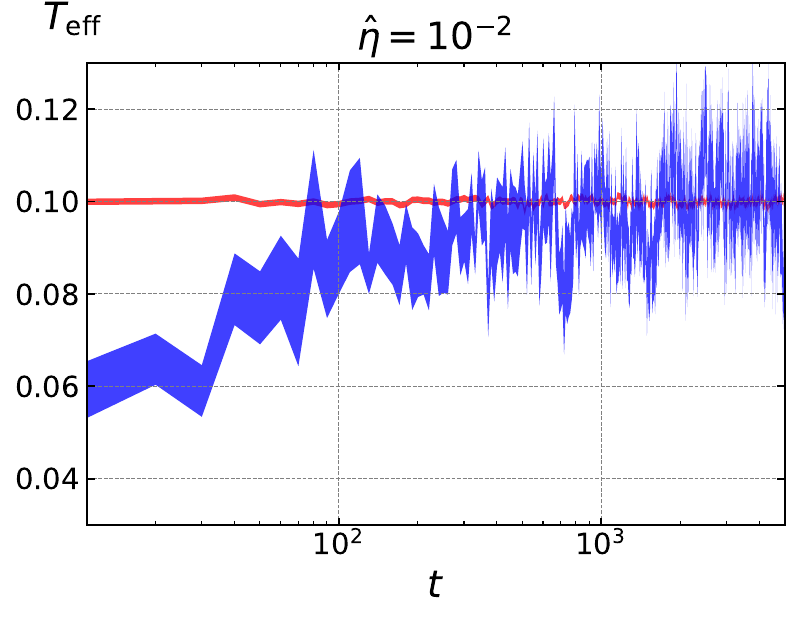}
	\end{minipage}
    \caption{Thermalization of long modes in the classical system described by the Langevin equation (\ref{LangEq}) with $\heta=10^{-3}$ ({\bf left}) and $\heta=10^{-2}$ ({\bf right}). 
    The Fourier modes with $k<1$ are initially prepared in a thermal state with the temperature $\Teff$ twice smaller than the temperature of the thermostat $\TT=0.1$. The blue (red) line shows $\Teff$ of the modes with $k<1$ ($k>1$). The width of the lines corresponds to the measurement uncertainty. }
	\label{fig:therm_Lang}
\end{figure}

We solve \cref{LangEq} numerically using a 3rd order stochastic, spectral, operator-splitting scheme~\cite{telatovich2017strong}, see Appendix~\ref{app:num} for details. We use the grid parameters listed in Sec.~\ref{ssec:units} and the time step $h=2.5\cdot 10^{-3}$.
First, we repeat the numerical exercise performed in Sec.~\ref{ssec:therm} to study the thermalization rate.
Namely, we take the lattice model (\ref{E_d}) with a positive sign of the self-interaction, 
prepare a suite of simulations in thermal equilibrium around the vacuum, 
remove half of the power from the long modes, and evolve them with the discretized version of \cref{LangEq}. Measuring the effective temperature of the long modes, we monitor
how quickly the heat bath restores the equilibrium.
The results are shown in Fig.~\ref{fig:therm_Lang} for the two values of the dissipation coefficient. They should be compared with Fig.~\ref{fig:therm}(b) (note the different time scale in the two figures) and confirm the validity of the estimate (\ref{t_therm_eta}).

\begin{figure}[t]
    \begin{minipage}[h]{0.49\linewidth}
        \centering
        \includegraphics[width=1.0\linewidth]{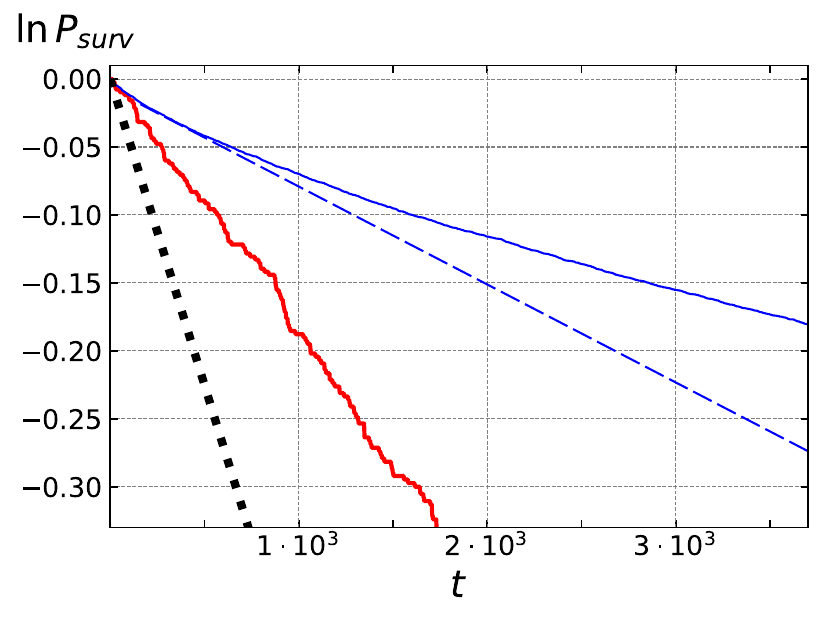} \\ (a)
	\end{minipage}
	\begin{minipage}[h]{0.49\linewidth}
        \centering
        \includegraphics[width=1.0\linewidth]{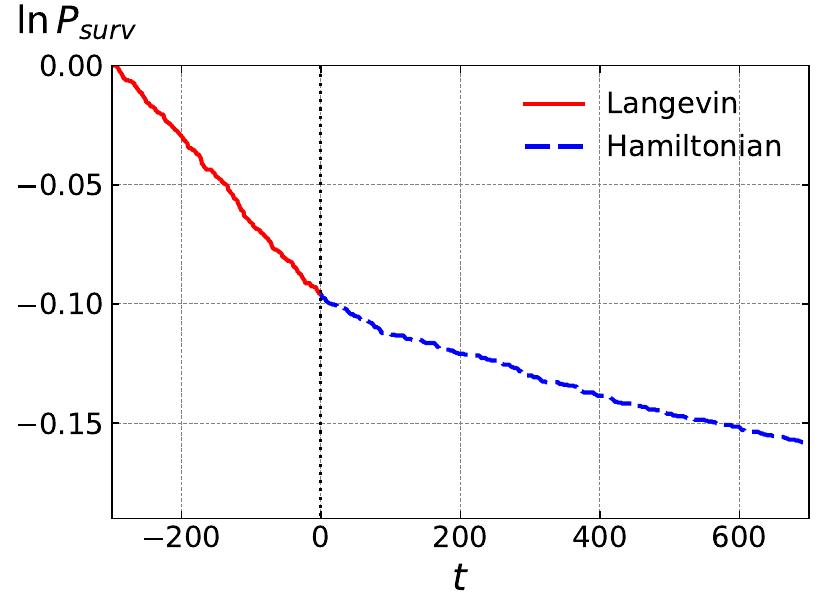} \\ (b)
	\end{minipage}
    \caption{\textbf{(a)} From top to bottom: survival probability in the Hamiltonian ensemble (thin blue), straight line tangent to it at $t\to 0$ (dashed blue), survival probability in the ensemble evolved using the Langevin equation with $\heta=10^{-2}$ (thick red), Euclidean theory prediction (dotted black). 
        \textbf{(b)}~Break in the survival probability curve upon transitioning from Langevin ($\heta=0.1$) to Hamiltonian ($\heta=0$) evolution at $t=0$. The temperature in all cases is $\TT=0.1$.}
	\label{fig:Psurv_Lang}
\end{figure}

Next, we return to the model with negative self-coupling. We prepare an ensemble of simulations in thermal equilibrium around the false vacuum and let it evolve according to the stochastic \cref{LangEq}.
Counting the surviving configurations at different moments of time we obtain a family of the survival probability curves $\ln P_{surv}(t)$ for different values of $\TT$ and $\heta$. A typical representative corresponding to $\TT=0.1$, $\heta=10^{-2}$ is shown in  
Fig.~\ref{fig:Psurv_Lang}(a) by the thick red line. It must be contrasted with the survival probability curve of the Hamiltonian ensemble ($\heta=0$) with the same temperature (thin blue line). We see that the curve with $\heta=10^{-2}$ does not exhibit any systematic flattening: it approximately follows a straight line, up to some random statistical fluctuations.\footnote{The simulation suits with the Langevin dynamics are smaller, and the respective statistical fluctuations are larger, than those for the Hamiltonian evolution. This is due to higher computational costs associated with the numerical solution of the stochastic \cref{LangEq}, compared to its Hamiltonian counterpart~(\ref{Hameqs}).}
This is expected, since now $t_{th}<t_{dec}$ and the coupling to the heat bath maintains the statistical equilibrium of the ensemble, counter-acting the classical Zeno effect.
We also observe that the slope of the line is steeper than for the Hamiltonian ensemble, but still less than the prediction of the Euclidean theory (black dotted line). 

To illustrate the difference between the dynamics of vacuum decay in the Hamiltonian and Langevin regimes, we perform the following numerical experiment.
We evolve an ensemble with \cref{LangEq} for $t\gg\heta^{-1}$, allowing all surviving configurations to reach equilibrium with the heat bath.
Then we abruptly decouple them from the thermostat by setting $\heta=0$.
As shown in Fig.~\ref{fig:Psurv_Lang}(b), the rate of decays changes sharply from the one corresponding to the Langevin evolution to the one corresponding to the Hamiltonian evolution. 
This confirms that the discrepancy (\ref{Aration}) between the Hamiltonian simulations and the Euclidean theory is not an artifact of some unknown systematics in the initial data or a subtle misidentification of the ensemble temperature. Indeed, the Langevin evolution erases any memory of the initial data if $t\gg\heta^{-1}$ and sets up the non-perturbative canonical ensemble with temperature $\TT$ appearing in the noise amplitude (\ref{Noise}). 

\subsection{Decay rate}

We turn to measuring the decay rate in the theory with the Langevin evolution.
We run a series of simulations covering the range $0.09\leqslant \TT\leqslant 0.13$, which is the same as in Sec.~\ref{sec:sim}, and $10^{-3}\leqslant\heta\leqslant 10$.
The range of $\heta$ includes the weak-damping regime where, depending on $\TT$, the thermalization time can be much longer or shorter than the average decay time.
In the first case, we split the probability curve into several segments and use the extrapolating procedure described in Sec.~\ref{ssec:unbiased}.
In the second case, the slope of $\ln\:P_{surv}$ is constant and we extract the rate directly using \cref{Psurv}. 

We find that the exponential part of the decay rate is insensitive to the details of the dynamics, so that $B$ in \cref{log_G(T)} is still given by the critical bubble energy $\Es$. Further exploring the decaying configurations with the methods described in Sec.~\ref{ssec:sph} we conclude that the critical bubble profile is unaffected by dissipation and noise.

\begin{figure}[t]
	\centering
    \includegraphics[width=0.6\linewidth]{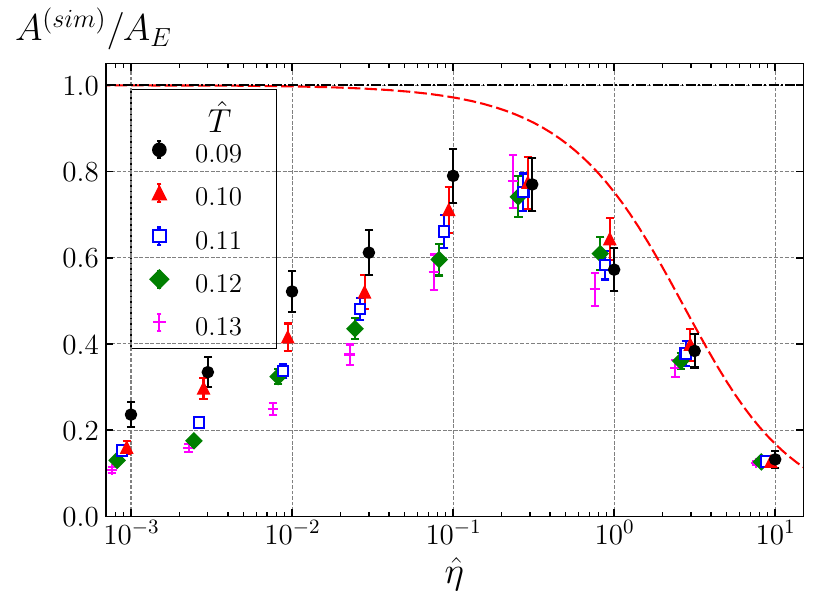}
    \caption{Prefactor of the decay rate measured in the Langevin simulations, normalized to the Euclidean prediction (\ref{G_E2}), as function of the dissipation coefficient $\heta$ for several values of temperature. The points corresponding to different temperatures are shifted in the horizontal direction for visualization purposes.
    The red dashed line is the prediction of the classical-statistical theory.}
	\label{fig:pref_Eta_temp}
\end{figure}

On the other hand, the prefactor exhibits a non-trivial dependence on $\heta$. Increasing $\heta$ from $10^{-3}$ to $0.1$ brings it closer to the Euclidean value, see Fig.~\ref{fig:pref_Eta_temp}. Thus, the Euclidean theory becomes more accurate for systems with more efficient thermalization. However, if dissipation is too strong, $\heta\gtrsim 1$, the long field modes become overdamped, which inhibits the decay. In this regime, the prefactor decreases again and therefore never reaches the Euclidean value.

The non-monotonic behavior of the prefactor in Fig.~\ref{fig:pref_Eta_temp} resembles the classical Kramers' result for the transition rate in one-dimensional stochastic systems~\cite{KRAMERS1940284} (see also~\cite{Hanggi:1990zz} for review). In field theory a similar dependence was observed in numerical simulations of the kink-antikink pair production~\cite{Alford:1991qg}. The high-$\heta$ tail of this curve is in fact universal~\cite{Langer:1969bc} and corresponds to the renormalization of the critical bubble growth rate $\omega_-$ due to dissipation, cf. eqs.~(\ref{GammaAff}) and (\ref{GammaLan}). This amounts to replacing the prefactor (\ref{G_E2}) with its classical-statistical version $A_{\rm stat}(\heta,\TT)$, such that
\be \label{pref_Langer}
\frac{A_{\rm stat}(\heta, \TT)}{A_E(\TT)} = \frac{1}{\hat{\o}_-}\left[\sqrt{\hat{\o}_-^2+\frac{\heta^2}{4}} -\frac{\heta}{2}\right] \;,
\ee
where $\hat{\o}_-=\sqrt{3}$, see \cref{omminus}. The dependence (\ref{pref_Langer}) is shown in Fig.~\ref{fig:pref_Eta_temp} with the red dashed line. Our simulation results agree well with it at $\heta\gtrsim1$. In particular, they approach the asymptotics $A^{(sim)}\propto \heta^{-1}$ at $\heta\gg 1$.

As we review in Appendix~\ref{app:Langer}, the derivation of Ref.~\cite{Langer:1969bc} breaks down at low dissipation. The condition for its validity in our case reads 
\be \label{etaLanger}
    {\heta}>\frac{\hat{\o}_- \TT}{\Es}\sim 0.15\;.
\ee
At smaller $\heta$ we indeed see a significant discrepancy between \cref{pref_Langer} and the data. At $\heta<10^{-3}$ we essentially recover the ratio (\ref{Aration}) measured in the Hamiltonian dynamics. Note that the situation differs from the one-dimensional mechanical case, where the classical transition rate completely vanishes at zero noise ($\eta=0$)~\cite{KRAMERS1940284,Hanggi:1990zz}. Note also that for fixed $\heta$ the condition (\ref{etaLanger}) will be always satisfied at low enough temperature. Hence, the measured prefactor $A^{(sim)}$ is expected to approach $A_{\rm stat}(\heta,\TT)$ from below at $\TT\to 0$ for any $\heta>0$. This is consistent with the trend exhibited by the simulation data.
In particular, we see from Fig.~\ref{fig:pref_Eta_temp} 
that the ratio $A^{(sim)}/A_E$ visibly increases in the range ${10^{-3}\lesssim \heta\lesssim 0.1}$ as the temperature gets lower. Still, at $\heta\sim10^{-3}$ 
it remains far from the limit
(\ref{pref_Langer}) even for $\TT=0.09$.

\section{Criterion for equilibrium decay rate}
\label{sec:criterion}

We have seen from the analysis of the Langevin dynamics that the statistical-mechanical expression (\ref{GammaLan}) for the decay rate, obtained under the assumption of thermal equilibrium, applies only if the dissipation coefficient is large enough, \cref{etaLanger}. It is suggestive to write this condition as an upper limit on the thermalization time using the relation (\ref{t_therm_eta}). Restoring the physical units we have, 
\be
\label{tthlimit}
t_{th}<\frac{E_b}{\omega_- T}\;.
\ee
In this form, the condition is applicable both to stochastic and Hamiltonian systems. We propose it as a general criterion for the validity of the equilibrium nucleation picture and the formulas (\ref{GammaAff}), (\ref{GammaLan}) for the vacuum decay rate. 

Given the expression (\ref{t_therm3}) for the thermalization time in the isolated model (\ref{S}), we readily see that the criterion (\ref{tthlimit}) is violated at all temperatures, whenever the theory is weakly coupled and the false vacuum decay is suppressed, $\TT < 1$. Notably, the violation of \cref{tthlimit} becomes stronger at lower temperatures. Thus, the Euclidean formula for the rate (\ref{GammaAff}) always fails in this model, no matter how low the temperature is.  

The latter property is not a peculiarity of the $(1+1)$ dimensional theory. The thermalization time typically grows faster than the right hand side of \cref{tthlimit} when the temperature decreases in any (classical) single-field theory. Consider, for example, $\lambda \phi^4$-theory in $(3+1)$ dimensions. The energy exchange between modes is now dominated by the $2\leftrightarrow 2$ scattering, and the reasoning similar to that carried in Sec.~\ref{ssec:therm} yields an estimate for the thermalization time,
\be
\label{ttherm3d}
t_{th}\sim \frac{(2\pi)^2 m}{\lambda^2 T^2}\;.
\ee
On the other hand, the critical bubble energy and its negative mode eigenvalue are 
\be
\label{Eb3d}
E_b\sim \frac{4\pi m}{\lambda}~,\qquad\omega_-\sim m\;.
\ee
We again see that the criterion (\ref{tthlimit}) is violated whenever $T<E_b$, i.e. whenever the false vacuum decay is suppressed. 

In theories with multiple fields and couplings the thermalization time can be shorter due to the interactions of the decaying field with other species. Of course, in this case, one also has to take into account the effect of extra species on the critical bubble free energy and growth rate. Thus, the criterion (\ref{tthlimit}) must be replaced with the condition (\ref{criterion}) presented in the Introduction. The latter may or may not be satisfied depending on the details of the interactions and must be checked on the case-by-case basis. 

It is worth stressing that the preceding discussion refers to systems at high temperature, in which the phase transition proceeds through classical nucleation of bubbles. The latter appear as {\it static} solutions in the Euclidean formalism. At low temperature, $T\lesssim m$, with $m$ being the mass of the decaying field, the false vacuum decay is described by {\it time-dependent} Euclidean solutions --- periodic instantons --- which correspond to quantum tunneling of the field through the potential barrier~\cite{Weinberg:2012pjx}. Our arguments comparing the time scales of thermalization and bubble nucleation in real time, as well as the criterion (\ref{criterion}), do not apply to this case. Testing the Euclidean derivation of the decay rate in the quantum regime is beyond the scope of this work.

\section{Oscillonic precursors}
\label{sec:osc}

In this section, we study the dynamics of the system before the bubble nucleation.
We find that the field evolution contains interesting features which further support the importance
of non-equilibrium processes in thermal false vacuum decay.

\begin{figure}[t]
    \centering
    \includegraphics[width=1.\linewidth]{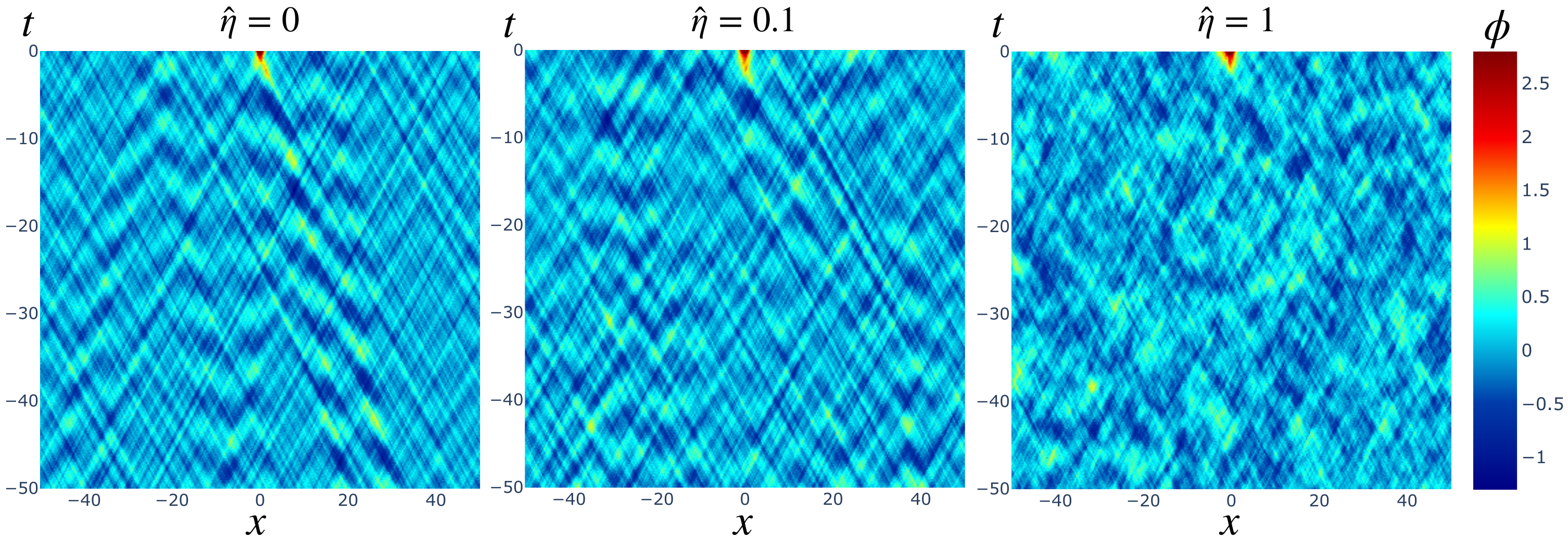}
    \caption{Typical evolution of the field preceding the critical bubble formation at $(t,x)=(0,0)$. We take $\TT=0.13$. {\bf Left:} In the case of no dissipation and noise ($\heta=0$) we see several non-linear waves --- oscillons --- propagating in the background of short, linear ultra-relativistic modes. One of the oscillons `collides' with a larger-amplitude thermodynamic fluctuation, which appears to trigger the decay event.  {\bf Middle:} In the case of moderate dissipation ($\heta=0.1$), the oscillonic precursor to the critical bubble is still visible but does not survive beyond one or two oscillation lengths. 
    {\bf Right:}~At large dissipation ($\heta=1$) there remain no long-wavelength propagating structures in the system.}
	\label{fig:2dsph}
\end{figure}

We run a series of simulations of the lattice $\l\phi^4$-theory and record the field history over a large time interval preceding the decay.
It is instructive to compare the histories obtained at different values of the dissipation coefficient.
Fig.~\ref{fig:2dsph} shows typical field profiles $\phi(x,t)$ for three values of $\heta$.
Recall that prior to the decay, most of the field's energy is contained in its freely-propagating, short, relativistic modes.
Apart from these linear waves, at $\heta=0$ we observe a population of slower-moving non-linear waves.
They are quasi-periodic and oscillate at low frequencies, $\hat{\o}_{\rm osc}<1$.
One of these waves eventually gives rise to the critical bubble.
At moderate dissipation, $\heta\sim 0.1$, the non-linear structures are still visible, but they decohere and dissipate over a time comparable to their oscillation period.
Finally, in the overdamped limit, $\heta\gtrsim 1$, the propagating non-linear structures disappear.

The non-linear waves, which populate the system at small dissipation and noise, are known as oscillons. 
They are localized, long-living solutions of classical equations of motion, arising in theories with anharmonic potentials~\cite{Amin:2010jq, Zhang:2020bec, Levkov:2022egq, Levkov:2023ncb}.
In cosmology, they have been extensively studied as a part of non-equilibrium processes involving scalar fields, such as the inflaton field, in the early universe~\cite{Johnson:2008se, Amin:2010xe, Amin:2010dc, Amin:2011hj, Gleiser:2011xj, Lozanov:2017hjm}.
Their role as `precursors' to the critical bubble in the context of false vacuum decay
has been investigated in Refs.~\cite{Gleiser:1991rf, Gleiser:1993pt, Gleiser:2004iy, Gleiser:2007ts, Pirvu:2023plk}.
Here we point out that oscillons are a part of non-equilibrium dynamics of the system, on its way from an initially thermal state to the critical bubble formation.
Their presence is a consequence of inefficient thermalization, $t_{th}\gg 2\pi/\hat{\o}_{\rm osc}$.
The strong coupling to the thermostat, which is necessary for the validity of the classical-statistical formalism, erases any such coherent objects.

To implement a quantitative test for the presence of the oscillonic precursor, we proceed as in Ref.~\cite{Pirvu:2023plk}. 
We first produce the envelope of each field configuration preceding the decay using the following procedure.
We remove the field average, so that $\ev{{\phi}(x,t)} = 0$ over the spacetime region of interest, e.g. the part of the simulation shown in Fig.~\ref{fig:2dsph}.
Next, we take the Fourier transform along the time dimension and 
set the amplitude of the negative-frequency Fourier modes to zero.
We then double the positive-frequency part and take its inverse Fourier transform. This gives a complex function $\Phi(x,t)$ whose real part coincides with the original field $\phi(x,t)$, whereas the imaginary part is the Hilbert transform of $\phi(x,t)$. The sought-after signal envelope is the absolute value $\left|\Phi(x,t)\right|$. Roughly speaking, this procedure removes the field oscillations, leaving their (spacetime dependent) amplitude.

\begin{figure}[t]
    \begin{minipage}[h]{0.32\linewidth}
        \centering
        \includegraphics[width=1.05\linewidth]{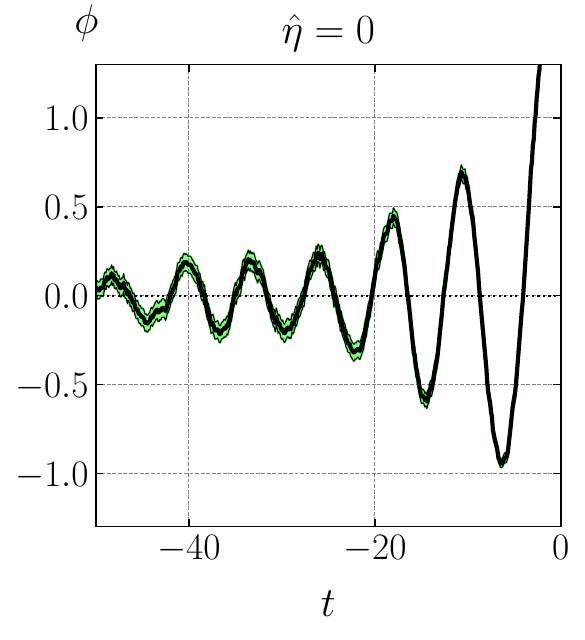} 
	\end{minipage}
	\begin{minipage}[h]{0.32\linewidth}
        \centering
        \includegraphics[width=1.05\linewidth]{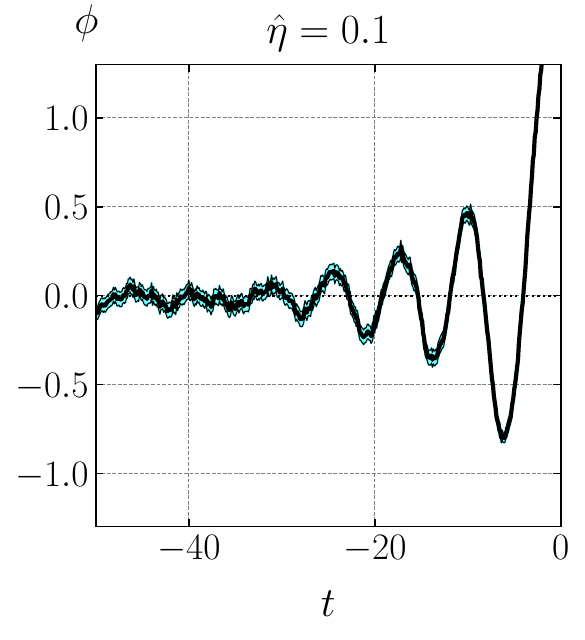} 
	\end{minipage}
	\begin{minipage}[h]{0.32\linewidth}
        \centering
        \includegraphics[width=1.05\linewidth]{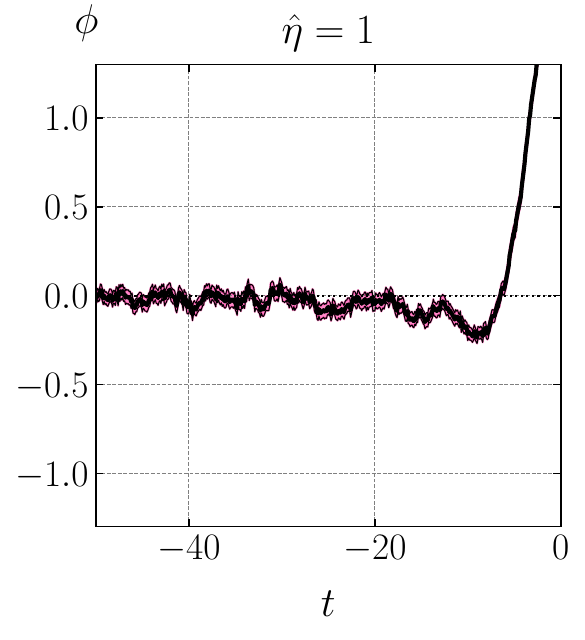} 
	\end{minipage} 
    \caption{The field along the trajectory of the oscillonic precursor to the critical bubble, at $\TT=0.13$ and different values of $\heta$. Each plot is the result of averaging over $200$ realizations.}
	\label{fig:oscs}
\end{figure}

Starting from the spacetime coordinates of each decay event, we trace backwards in time the position of the maximal peak of the envelope $\left|\Phi(x,t)\right|$. This produces the trajectory of the peak which we denote as $x_{\rm osc}(t)$.
For each simulation, we take the field value ${\phi}\big(x_{\rm osc}(t), t\big)$ along the resulting trajectory. 
We stack these functions from different simulations with respect to the reference nucleation time at $t=0$.
Examples of the average precursors resulting from this procedure are shown in Fig.~\ref{fig:oscs} for three values of the dissipation coefficient $\heta$.
The lifetime of the oscillon is clearly reduced when $\heta$ increases. In other words, the oscillon precursors are present if the thermalization time in the system is long enough, which also corresponds to the failure of the equilibrium expression for the bubble nucleation rate. We leave a further exploration of the connection between the oscillonic precursors and the vacuum decay rate for future.

The process of vacuum decay with the oscillonic precursor bears some similarities with the flyover transitions studied in Ref.~\cite{Blanco-Pillado:2019xny}.
In both cases the system contains a coherent fluctuation of the field with non-zero time derivative that leads to the decay.
An important difference, however, is that in our case this fluctuation is not present in the initial conditions, but 
emerges as a result of dynamical evolution. The dynamics picks up the most probable decay path culminating in the formation of the critical bubble. Consequently, 
the exponential part of the decay rate coincides with the Boltzmann suppression of the critical bubble.

\section{Discussion and outlook}
\label{sec:concl}

In this paper we studied the dynamics of a real classical scalar field with unstable potential in $(1+1)$ dimensions. Our goal was to test the predictions of the Euclidean formalism for the false vacuum decay at finite temperature. While we confirmed some of the predictions, such as the shape of the critical bubble and the leading exponential suppression of the decay rate, we found a disagreement, by almost an order of magnitude, in the value of the decay rate prefactor. We identified the cause of this discrepancy with the violation of thermal equilibrium during the bubble nucleation process. Such violation is not an exceptional property of our setup, but rather a general rule. In particular, it is unavoidable in weakly coupled single-field theories. Whether or not it occurs in multi-field models must be verified case by case. For this purpose, we proposed a quantitative criterion setting an upper bound on the thermalization time in the system that must be satisfied in order for the equilibrium theory to work. It will be interesting to test this criterion by numerical simulations of false vacuum decay in different models. We plan to carry out such research in the future.

One may wonder if the Euclidean results for the decay are of any use when the equilibrium condition does not hold. By analogy with stochastic mechanical systems~\cite{Hanggi:1990zz}, it is tempting to conjecture that in this case the equilibrium theory provides an {\it upper bound} on the decay rate. This seems to be supported by our findings about the $\lambda\phi^4$-theory where indeed the rate is lower than the Euclidean prediction. However, generalizing this conclusion to more complicated theories would be premature. In the presence of multiple fields, the height of the barrier between the false and true vacua, which determines the exponential suppression of the decay rate, is significantly altered by excitations. The Euclidean theory computes this height as the effective free energy of the critical bubble, assuming that all fields are in thermal equilibrium. This calculation becomes moot if the equilibrium condition breaks down and it is unclear a priori how the barrier height will be affected. In principle, it is possible that in the models where the bubble free energy is higher than its bare energy, the lack of thermal equilibrium may lower the barrier, leading to enhancement of the decay rate. Clearly, the dynamics of false vacuum decay in multi-field models deserves a thorough study. 

Examples where non-equilibrium dynamics affects the exponential suppression of the decay rate have been known in the literature. They include bubble nucleation after a quench~\cite{Gleiser:2004iy, Gleiser:2007ts}, phase transitions at preheating~\cite{Kofman:1995fi, Khlebnikov:1998sz}, and catalysis of false vacuum decay by black holes~\cite{Hiscock:1987hn, Berezin:1990qs, Gregory:2013hja, Tetradis:2016vqb, Gorbunov:2017fhq, Mukaida:2017bgd, Kohri:2017ybt, Hayashi:2020ocn, Shkerin:2021zbf, Shkerin:2021rhy, Strumia:2022jil, Briaud:2022few}. Our present results show that the thermal false vacuum decay in flat spacetime is not fundamentally different from these processes. They strongly motivate development of theoretical techniques, beyond the direct numerical simulations, that would not rely on thermal equilibrium to describe phase transitions. This is particularly important in the regime of strongly suppressed transitions where direct real-time simulations are unfeasible.   

One option here is the hybrid method of Refs.~\cite{Moore:2000jw, Moore:2001vf, Gould:2022ran} which combines Euclidean Monte-Carlo sampling with real-time evolution of configurations containing the critical bubble. It will be interesting to understand if this method captures the non-equilibrium effects uncovered in this work. Intriguingly, applying this method to a $(3+1)$-dimensional scalar field model, 
Ref.~\cite{Gould:2024chm} recently found a significantly lower phase transition rate than predicted by the Euclidean theory. It remains to be seen if this discrepancy can be due to the non-equilibrium dynamics of bubble nucleation.

Another promising direction is provided by a class of semi-classical methods devised for the description of dynamical tunneling in various non-equilibrium systems~\cite{miller1974classical, Rubakov:1992ec, Kuznetsov:1997az, Bonini:1999kj, Bezrukov:2003er, Bezrukov:2003tg, Levkov:2004ij, Levkov:2007yn, Levkov:2008csa, Demidov:2011dk, Demidov:2015bua, Demidov:2015nea}. This approach can, in principle, allow us to go beyond the high-temperature classical approximation and address the full quantum dynamics of the false vacuum decay. It is worth reminding at this point that our criterion of thermal equilibrium (\ref{criterion}) has been derived within classical theory and its applicability in the quantum regime is not known at present.
Validity of the Euclidean theory of false vacuum decay at finite temperature in the quantum regime is an interesting open question. We leave its exploration for future.

Finally, we note that our results concerning the deviation of the transition rate from the Euclidean prediction may not apply to non-perturbative processes other than the false vacuum decay. For example, the sphaleron transitions leading to violation of fermion numbers in the Standard Model~\cite{Kuzmin:1985mm, Rubakov:1996vz} can be in detailed balance in thermal plasma. This is an important difference from the false vacuum decay which is a one-way process. Indeed, the rate of sphaleron transitions measured in real-time simulations of simplified toy models agrees well with the Euclidean results~\cite{Grigoriev:1989je, Grigoriev:1989ub}. Still, our findings suggest that sphaleron transitions, at least in the regime when they are exponentially suppressed, may also possess non-equilibrium features. Thus, a scrutiny of their real-time dynamics is warranted.

\section*{Acknowledgments}

We thank Mustafa Amin, Asimina Arvanitaki, Cliff Burgess, Claudia Cornella, Marco Costa, Ruth Gregory, Junwu Huang, Matthew Johnson, Alexander Kayssi, Juraj Klaric, Andrew Kovachik, Sung-Sik Lee, Duncan O'Dell, Alexander Penin, Maxim Pospelov, Jury Radkovski, Kam To Billy Sievers, Mikhail Shaposhnikov and Andrei Zelnikov for fruitful discussions. We are grateful to Oliver Gould for instructive correspondence.
Research at Perimeter Institute is supported in part by the Government
of Canada through the Department of Innovation, Science and Economic
Development Canada and by the Province of Ontario through the Ministry
of Colleges and Universities. The work of SS is supported by the Natural Sciences and Engineering Research Council (NSERC) of Canada. This research was enabled in part by support provided by
Compute Ontario (www.computeontario.ca) and Digital Research Alliance of Canada
(alliancecan.ca).

\appendix

\section{Revisiting classical-statistical transition theory}
\label{app:Langer}

In this Appendix we review the derivation of the false vacuum decay rate in classical multi-dimensional stochastic systems. We follow Ref.~\cite{Langer:1969bc} (see also~\cite{Hanggi:1990zz}).

Consider a system described by coordinates $q_i$, $i=1, \dots ,N$, and momenta $p_i$, with the Hamiltonian
\be
\label{appHam}
H=\sum_i\frac{p_i^2}{2} + U(q)\;.
\ee
The potential has a local minimum at $q=0$ where we set $U(0)=0$. This minimum is separated by a potential barrier from the region with negative potential energy. We denote the saddle point of the potential by $q^{(b)}$ and the corresponding energy by $E_b\equiv U(q^{(b)})$. 

The system is coupled to an external heat bath with temperature $T$, which introduces dissipation and stochasticity in its dynamics. For simplicity, we assume that the dissipation is linear in momenta and the dissipation coefficient $\eta$ is the same for all degrees of freedom. The equation of motion takes the form, 
\be
\label{appeoms}
\dot q_i=p_i\;,\qquad \dot p_i=-\frac{\d U}{\d q_i} -\eta p_i+\xi_i\;,
\ee
where $\xi_i$ is a white noise obeying the classical fluctuation-dissipation theorem,
\be
\label{appnoise}
\ev{\xi_i(t)\xi_j(t^\prime)}=2\eta T\,\delta_{ij}\,\delta(t-t^\prime)\;. 
\ee

Next, we introduce the phase-space distribution function ${\cal P}(z)$, where $z_I$, $I=1, \dots ,2N$, collectively denote the phase-space coordinates, $z=(q_i,\dots,q_N,p_1,\dots,p_N)$. It follows from eqs.~(\ref{appeoms}) that ${\cal P}(z)$ obeys the Fokker--Planck equation,
\be \label{FP}
\frac{\d {\cal P}}{\d t} = -\sum_I \frac{\d {\cal J}_I}{\d z_I}~,\qquad
{\cal J}_I=-\sum_{J} M_{IJ}\left( \frac{\d H}{\d z_J} + T\frac{\d }{\d z_J} \right){\cal P}\;, 
\ee
with the matrix $M_{IJ}$ of the block form
\be
\label{appM}
M=\begin{pmatrix} 
   0 & -\mathbbm{1} \\
   \mathbbm{1}  & \eta\mathbbm{1}
\end{pmatrix}\;,
\ee
where $\mathbbm{1}$ stands for the unit $N\times N$ matrix.
We are interested in the stationary solution to this equation that describes the current of the density of states across the barrier. We search for it in the form,
\be \label{app:p}
{\cal P}(z)={\cal P}_{eq}(z)\zeta(z) \;, \qquad {\cal P}_{eq}(z)=\frac{1}{Z}\e^{-H(z)/T}\;.
\ee 
Here ${\cal P}_{eq}$ is the equilibrium distribution normalized to the partition function of the false vacuum $Z$, and $\zeta(z)$ is $1$ in the neighborhood of the false vacuum and $0$ in the region outside the barrier. A key assumption is that $\zeta(z)$ goes between these two asymptotics in a small vicinity of the saddle point, where the Hamiltonian can be expanded to quadratic order,
\be \label{H_exp}
H \approx E_b +\frac{1}{2}\sum_{I,J}{\cal E}_{IJ}\left(z_I-z_I^{(b)}\right)\left(z_J-z_J^{(b)}\right) \;,\qquad {\cal E}=\begin{pmatrix}
   \frac{\d^2 U}{\d q^2}  & 0 \\
   0  & \mathbbm{1}
\end{pmatrix} \;.
\ee 
We will return to the validity of this assumption below.

Substituting the expressions (\ref{app:p}), (\ref{H_exp}) into \cref{FP} we obtain,
\be \label{FP2}
\sum_{I,J}M_{IJ}\left(- \sum_K {\cal E}_{IK}\left(z_K-z_K^{(b)}\right)+T\frac{\d }{\d z_I}\right)\frac{\d \zeta}{\d z_J} = 0\;.
\ee 
We further assume that the function $\zeta(z)$ depends on a single linear combination of the phase-space coordinates. Inserting the Ansatz
\be \label{f(u)}
\zeta(z) = f(w) \;,\qquad w = \sum_I W_I \left(z_I-z_I^{(b)}\right) \;
\ee 
into \cref{FP2} we see that it goes through if $W_I$ is an eigenvector of the matrix ${\cal E}M$,
\be \label{ev_problem}
\sum_{I,J}{\cal E}_{KI} M_{IJ}\, W_J = -\kappa\, W_K\;.  
\ee 
Normalizing it as 
\be \label{norm2}
\sum_{I,J}W_I M_{IJ} W_J =1\;, 
\ee
we obtain the equation for the function $f(w)$,
\be \label{f(u)eq}
T f^{\prime\prime}(w) + \kappa w f^\prime(w) = 0 \;.
\ee 
If $\kappa>0$ this equation has a solution with the asymptotics $f\to 0$ at $w\to +\infty$ and $f\to 1$ at $w\to -\infty$, which reads,  
\be \label{f(u)sol}
f(w)=\sqrt{\frac{\kappa}{2\pi T}}\limitint_w^{\infty} \diff w^\prime \exp\left(-\frac{\kappa w^{\prime \, 2}}{2 T}\right) \;.
\ee 
The surface $w=0$ passes through the saddle point of the potential and separates the false vacuum at $w<0$ from the true vacuum at $w>0$.

For the choice of the matrices $M$ and ${\cal E}$ in eqs.~(\ref{appM}), (\ref{H_exp}) we can give a more detailed expression for the vector $W_I$. Splitting it as
\be
\label{appWsplit}
W=\begin{pmatrix}
W^{(1)}\\
W^{(2)}
\end{pmatrix}
\ee
and inserting this form into eqs.~(\ref{ev_problem}), (\ref{norm2}) we obtain,
\be
\label{appWsplit2}
W^{(1)}_i=\frac{\kappa+\eta}{\sqrt{\eta}} \bar{W}_i~\;, \qquad W^{(2)}_i=-\frac{1}{\sqrt{\eta}} \bar{W}_i\;,
\qquad i=1, \dots , N\;.
\ee
Here $\bar{W}$ is the negative eigenvector of the potential at the saddle point,
\be
\label{appbarW}
\left(\frac{\d^2 U}{\d q^2}\right)_{ij}\bar{W}_j=-\omega_-^2 \bar{W}_i\;,
\ee
which we normalize to unity, $\sum_i \bar{W}_i^2=1$. Note that the expressions (\ref{appWsplit2}) make sense only if the dissipation coefficient $\eta$ is non-vanishing. In addition, we obtain an equation for the eigenvalue,
\be
\label{appkappa}
\kappa (\kappa+\eta)=\omega_-^2\quad \Longrightarrow\quad \kappa=\sqrt{\omega_-^2+\frac{\eta^2}{4}}-\frac{\eta}{2}\;,
\ee
where we have picked the positive root.

The decay rate is given by the probability flux through the surface $w=0$. The probability current on the solution reads,
\be
\label{appJ}
{\cal J}=\frac{1}{Z}\sqrt{\frac{\kappa}{2\pi\eta T}}\begin{pmatrix}
\bar{W} \\ \kappa\bar{W}\end{pmatrix}
\exp\left[-\frac{1}{T}\left(H+\frac{\kappa w^2}{2}\right)\right]\;.
\ee
It is convenient to separate the coordinate and momentum $q_-$, $p_-$ corresponding to the negative mode, and collectively denote the rest of the phase-space variables by $z_+$. Then we have
\be
\label{appwexpl}
w=\frac{1}{\sqrt\eta}\left[(\kappa+\eta)\big(q_--q_-^{(b)}\big)-p_-\right]\;,
\qquad H=E_b+\frac{p_-^2}{2}-\frac{\omega_-^2\big(q_--q_-^{(b)}\big)^2}{2} + H_+(z_+)\;.
\ee
Computing the component of the current ${\cal J}_\perp$ orthogonal to the surface $w=0$ we obtain,
\be
\label{appGammaInt}
\Gamma=\limitint_{w=0} \diff z\, {\cal J}_\perp
=\frac{\e^{-E_b/T}}{Z} \int \diff q_-\sqrt{\frac{\kappa\eta}{2\pi T}}
\exp\left[-\frac{\eta(\kappa+\eta)}{2T}\big(q_--q_-^{(b)}\big)^2\right]
\cdot\int \diff z_+\,\e^{-H(z_+)/T}\;.
\ee
The integral over positive modes is Gaussian and produces the determinant of small stable fluctuations around the bubble. The same determinant arises in the Euclidean calculation of the critical bubble contribution into the free energy. In more detail,
\be
\label{appImF}
\Im F=T\frac{\e^{-E_b/T}}{Z}\cdot\frac{\pi}{\omega_-}\int \diff z_+\,\e^{-H(z_+)/T}\;.
\ee
Taking the Gaussian integral over $q_-$ in \cref{appGammaInt} explicitly and using \cref{appkappa}, we arrive at the 
final answer for the decay rate,
\be
\label{appGamma2}
\Gamma=\frac{\kappa}{\pi T} \Im F\;.
\ee
This is equivalent to \cref{GammaLan} from the Introduction. 

Let us now return to the assumption that the function $\zeta(z)$ in \cref{app:p} must interpolate between $1$ and $0$ in a small vicinity of the barrier. The Ansatz (\ref{f(u)}) clearly contradicts this assumption: the interpolation region extends along an infinite plane $w=0$. This is not a problem as long as the distribution function becomes quickly suppressed along this plane away from the saddle point, so that the far-off regions do not contribute into the probability flux. Let us work out the corresponding condition. From \cref{appGammaInt} we see that the integral over $q_-$ converges at 
\be
\label{appqestim}
\left|q_--q_-^{(b)}\right| \sim \sqrt{\frac{T}{\eta(\kappa+\eta)}}\;.
\ee
The change of the potential energy at this distance must be smaller than $E_b$ --- otherwise the quadratic expansion (\ref{H_exp}) would break down. This yields the condition:
\be
\label{appcond}
\frac{\omega_-^2T}{\eta(\kappa+\eta)}\ll E_b\qquad\Longrightarrow\qquad\eta\gg\frac{\omega_- T}{E_b}\;,
\ee
where we have approximated $(\kappa+\eta)\approx \omega_-$ for $\eta\lesssim \omega_-$. 
We conclude that the applicability of the solution (\ref{app:p}), (\ref{f(u)}), (\ref{f(u)sol}) implies a lower bound on the dissipation coefficient.

The above analysis can be generalized to field theory.
In the case of a theory with multiple field species, the critical bubble energy in \cref{appcond} must be replaced with its effective free energy, leading to the condition (\ref{etaLan}) from the main text.

\section{Euclidean analysis of $\lambda\phi^4$ theory in 2d}
\label{app:det}

Here we compute several equilibrium observables in the theory (\ref{S}) using the Euclidean techniques. Namely, we compute the 1-loop effective potential used in Sec.~\ref{ssec:effpot} to extract the thermal mass correction. Then we derive the Euclidean prediction for the vacuum decay rate, including the prefactor. 
For completeness, we perform the full quantum computation, and take the classical limit in the end.
This allows us to see explicitly how the classical limit is achieved, and how the quantum and classical contributions factorize.

\subsection{Renormalization of the effective potential}
\label{app:det:mass}

We start with \cref{V-1-loop-int}. Focusing on the second term representing the 1-loop correction and taking the integral over momentum we get, 
\be \label{V2}
V^\text{1-loop}(\bar{\phi}) 
=\frac{T}{2}\sum_{n=-\infty}^\infty 
\left(\sqrt{\omega_n^2+m^2(\bar{\phi})}- \left| \omega_n \right|\right) \;.
\ee
For large Matsubara frequencies the terms in the sum behave as $m^2(\bar{\phi})/(2 \left| \omega_n \right|)$, so the sum itself logarithmically diverges. 

To renormalize this divergence, we consider the zero-temperature limit. In this case the sum becomes an integral,
\be
\label{Vvac1}
V^\text{1-loop}_{T=0}(\bar{\phi})=\int \frac{d\omega}{2\pi}\,
\left(\sqrt{\omega^2+m^2(\bar{\phi})}-\omega\right) 
=\frac{m^2(\bar{\phi})}{4\pi}\left(\ln\frac{2\Lambda}{m(\phi)}+\frac{1}{2}\right)\;,
\ee
where we have used a simple cutoff regularization $\omega<\Lambda$ to remove the divergence. 
Let us expand this expression at small $\bar{\phi}$:
\be
\label{Vvac3}
V^\text{1-loop}_{T=0}(\bar{\phi})=\frac{m^2}{4\pi}
\left(\frac{1}{2}+\ln\frac{2\Lambda}{m}\right)
-\bar{\phi}^2\frac{3\lambda}{4\pi}\ln\frac{2\Lambda}{m}+ \dots \; . 
\ee
The first term is removed by subtracting the zero-point energy, whereas the second term is absorbed by renormalizing the mass in the tree-level potential,
\be \label{vacrenorm}
m^2\mapsto m_{\rm ren}^2=m^2-\frac{3\lambda}{2\pi}\ln\frac{2\Lambda}{m}\;.
\ee
This is sufficient to render the renormalized effective potential finite. 
The eventual 1-loop contribution reads,
\be
\label{Vvac4}
V^\text{1-loop}_{T=0,\text{ren}}(\bar{\phi})=-\frac{3\lambda\bar{\phi}^2}{8\pi}
-\frac{m^2-3\lambda\bar{\phi}^2}{8\pi}\ln\left(1
-\frac{3\lambda\bar{\phi}^2}{m^2}\right) \;.  
\ee
Note that this expression is real only at $\bar{\phi}<m/\sqrt{3\lambda}$.

Returning to the case of general temperature (\ref{V2}), we renormalize the sum by the same subtraction as at $T=0$. This yields, 
\begin{align}
\label{V3}
V^\text{1-loop}_{\text{ren}}(\bar{\phi})&=\frac{Tm(\bar{\phi})}{2}+T\sum_{n=1}^\infty
\Big[\sqrt{\omega_n^2+m^2(\bar{\phi})}-\omega_n\Big]
-\frac{m^2}{8\pi}-\frac{m^2(\bar{\phi})}{4\pi}\ln\frac{2\Lambda}{m}\notag \\
&=\frac{Tm(\bar{\phi})}{2}
+T\sum_{n=1}^\infty\Bigg[\sqrt{\omega_n^2+m^2(\bar{\phi})}-\omega_n-\frac{m^2(\bar{\phi})}{2\omega_n}\Bigg]
\!+ \frac{m^2(\bar{\phi})}{4\pi}\Bigg[\ln\frac{m}{4\pi T}+ \gamma\Bigg]
\!-\!\frac{m^2}{8\pi}\;,
\end{align} 
where $\gamma$ is the Euler constant. 
In passing to the second line we have used 
\be
\label{harmsum}
\sum_{n=1}^{n_{\rm max}} \frac{1}{n}=\ln n_{\rm max}+ \gamma+O(1/n_{\rm
max})
\ee
and identified
\be
\label{omegacut}
\Lambda=2\pi T n_{\rm max}\;.
\ee 
The expression (\ref{V3}) is finite at any temperature and reduces to \cref{Vvac4} at $T\to 0$. In the opposite limit $T\gg m$ the first term in it dominates and we arrive at \cref{V-1-loop-reg} from the main text.

\subsection{Imaginary part of free energy}
\label{app:det:Mats}

The free energy is computed from the partition function (\ref{Zgen}).
The critical bubble is a saddle point of the path integral in $Z$, hence the partition function has a contribution
\be
\label{Zsph}
Z\ni 2\cdot\exp\left\lbrace -S_E[\phi_b] -\frac{1}{2}\ln{\cal D} + \frac{1}{2}\ln{\cal D}^{(0)} \right\rbrace\;,
\ee
where $S_E[\phi_b]=E_{b}/T$ is the critical bubble action, and ${\cal D}$, ${\cal D}^{(0)}$ are the determinants of small fluctuations around the bubble and the false vacuum, respectively.
The factor $2$ comes from the fact that the false vacuum can decay both to the left and to the right in the $\phi$-space. In other words, there are two nontrivial saddle points with field profiles $\pm \phi_b(x)$. 
It is well-known that ${\cal D}$ is negative, hence the right-hand side of \cref{Zsph} is imaginary.
This gives an imaginary contribution to the free energy of the system
\be
\label{ImF}
\Im
F=T\cdot2\cdot\frac{1}{2} \left| \frac{\D}{\D^{(0)}} \right|^{-1/2} \e^{-E_b/T}\;,  
\ee
where we included the factor $1/2$ coming from the integration over the negative mode~\cite{Coleman:1978ae}. 
The imaginary free energy (\ref{ImF}) is directly related to the Euclidean prediction for the false vacuum decay rate, see \cref{GammaAff}. 

The determinant ${\cal D}$ in \eqref{ImF} is the product over eigenvalues $\a_I$ of the equation
\be
\label{eigen1}
-\Box\vf_I+ \big(m^2-3\l\phi_b^2(x) \big) \vf_I=\a_I\, \vf_I\;.
\ee
Thermal nature of the partition function (\ref{Zgen}) implies periodic boundary conditions in the Euclidean time. 
Since the critical bubble is time-independent, we can use separation of variables:
\be
\label{separ}
\vf_I(x,\tau)=\e^{-i\omega_n \tau}\vf_{n,k}(x) \;,
\ee
where $\omega_n$ is the $n$th Matsubara frequency and the spatial functions satisfy
\be
\label{eigen2}
-\vf_{n,k}^{\prime\prime}+ \left(\omega_n^2+m^2-\frac{6m^2}{\ch^2{mx}}\right) \vf_{n,k}=
\alpha_{n,k}\, \vf_{n,k}\;. 
\ee
The determinant thus factorizes into a product over different Matsubara sectors,
\be
\label{factor}
\frac{\D}{\D^{(0)}}=\prod_{n=-\infty}^\infty \frac{\D_n}{\D^{(0)}_n}
=\frac{\D_0}{\D^{(0)}_0}\prod_{n=1}^\infty
\left(\frac{\D_n}{\D^{(0)}_n}\right)^2\;. 
\ee

Consider a pair of operators
\be
\label{pairO}
{\cal O}_\mu=-\d_x^2+ \mu^2-\frac{6m^2}{\ch^2{mx}}\;, \qquad
{\cal O}_\mu^{(0)}=-\d_x^2+ \mu^2\;,
\ee
with a general parameter $\mu$. 
One can evaluate the ratio of their determinants using the formula~\cite{Coleman:1978ae},
\be
\label{detform}
\frac{\D_\mu}{\D_\mu^{(0)}}=
\lim_{x\to+ \infty}\frac{\vf_\mu(x)}{\vf_\mu^{(0)}(x)}\;,
\ee     
where $\vf_\mu$, $\vf_\mu^{(0)}$ are solutions of the respective
equations with the same vanishing asymptotics at negative infinity,
\be
\label{Oeqs}
{\cal O}_\mu\vf_\mu=0 \; , \qquad
{\cal O}^{(0)}_\mu\vf_\mu^{(0)}=0 \; , \qquad
\vf_\mu=\vf_\mu^{(0)}= \e^{\mu x} \; , \quad x \to -\infty\;.
\ee
In fact, $\vf_{\mu}^{(0)}$ has the simple exponential form at all $x$,
so we only need to find $\vf_\mu$. The solution with the required
properties has the form,
\be \label{solphimu}
\vf_\mu(x)=c\, {\rm P}_2^\mu(\th x) \;, \quad c=-\mu(1-\mu)(2-\mu) \Gamma(-2-\mu) \;,
\ee
where ${\rm P}_2^\mu$ is the associated Legendre polynomial. 
The asymptotics of (\ref{solphimu}) at positive infinity is
\be \label{solphias}
\vf_\mu(x) \simeq c\, \frac{\e^{\mu x}}{\Gamma(1-\mu)} \; , \quad x \to + \infty\;.
\ee
Substituting it into \eqref{detform}, we obtain
\be \label{fracdet}
\frac{\D_\mu}{\D_\mu^{(0)}}=\frac{(\mu-m)(\mu-2m)}{(\mu+m)(\mu+2m)} \;.
\ee

The contribution of the non-zero Matsubara sectors into the prefactor in \cref{ImF} becomes
\be \label{prodnz}
\prod_{n=1}^\infty
\frac{\D_n^{(0)}}{\D_n}=
\prod_{n=1}^\infty
\frac{\big(\sqrt{\omega_n^2+m^2}+m\big) \big(\sqrt{\omega_n^2+m^2}+2m\big)}
{\big(\sqrt{\omega_n^2+m^2}-m\big) \big(\sqrt{\omega_n^2+m^2}-2m\big)}\;. 
\ee 
Note that it is positive as long as $\omega_1^2>3m^2$ or, equivalently, $T>T_c$, where
\be \label{Tsph}
T_c=\frac{\sqrt{3}}{2\pi}m \;.
\ee
Then a negative mode exists only in the $0$'th sector. It is straightforward to find its shape and eigenvalue from \cref{eigen2} with $\omega_n=0$. We have,
\be \label{alphaminus}
\vf_{0,0}\propto\frac{1}{\ch^2 mx}\;,\qquad\a_{0,0}\equiv-\omega_-^2=-3m^2\;.
\ee
On the other hand, if $T<T_c$, then the determinant $\D_1$ becomes negative implying that the critical bubble acquires an additional negative mode in the first Matsubara sector. 
This means that it does not dominate the vacuum decay anymore. 
The relevant solutions at these low temperatures are periodic instantons, and the decay proceeds via quantum tunneling.

\begin{figure}[tb]
    \centering
    \includegraphics[scale=0.6]{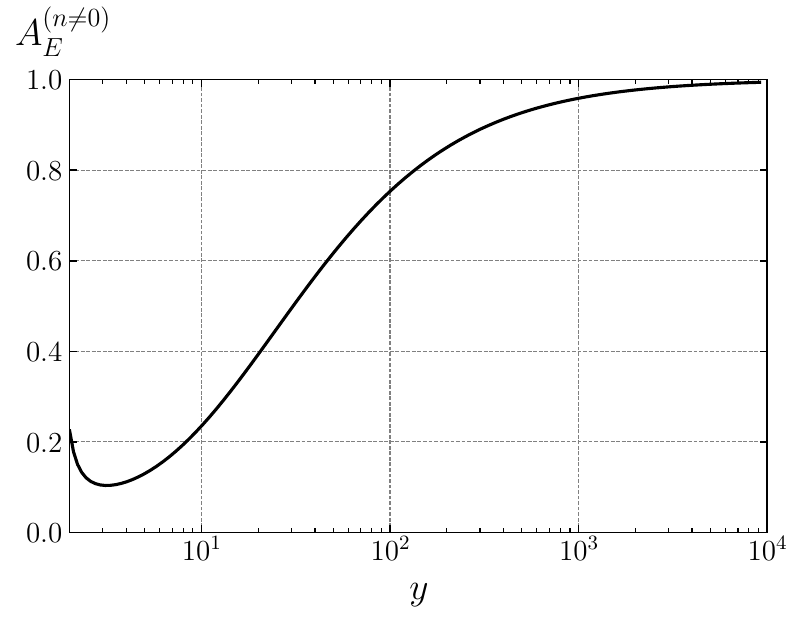}
    \caption{Dependence of the prefactor $A_E^{(n\neq 0)}$ due to non-zero Matsubara sectors on the ratio $y=2\pi T/m$. }
    \label{fig:Ahighn}
\end{figure}

The product (\ref{prodnz}) diverges at large $n$. In the imaginary part of the free energy (\ref{ImF})
this divergence is absorbed by the mass renormalization (\ref{vacrenorm}) in the leading exponent (recall \cref{Esph} for the critical bubble energy),
\be \label{exprenrom}
\exp\left\{-\frac{4m^3}{3\l T}\right\}\simeq
\exp\left\{-\frac{4m^3_{\rm ren}}{3\l T}\right\}
\exp\left\{-\frac{3m}{\pi T}\ln\frac{2\Lambda}{m}\right\} \;.
\ee 
We combine the second factor with (\ref{prodnz}), use the identification (\ref{omegacut}), and expand $\ln n_{\rm max}$ into the harmonic series using (\ref{harmsum}).
This gives the renormalized prefactor of the decay rate due to non-zero Matsubara sectors,
\be \label{Anhigher}
A_E^{(n\neq 0)}=\exp\left\{\frac{3m}{\pi T}\left(\!\gamma\!-\!\ln\frac{4\pi
  T}{m}\right) \right\}
\prod_{n=1}^\infty
\frac{\big(\sqrt{\omega_n^2\!+ \!m^2}\!+ \!m\big)
\big(\sqrt{\omega_n^2\!+ \!m^2}\!+ \!2m\big)}
{\big(\sqrt{\omega_n^2\!+ \!m^2}\!-\!m\big)
\big(\sqrt{\omega_n^2\!+ \!m^2}\!-\!2m\big)}\, 
\e^{-6m/\omega_n}\;.
\ee
It depends on the ratio $y\equiv 2\pi T/m$. This dependence is shown in Fig.~\ref{fig:Ahighn}. 
We observe that $A^{(n\neq 0)}_E$ varies between $0.1$ at $y\sim 1$ and $1$ at $y\to \infty$. The latter property shows that non-zero Matsubara modes do not affect the decay rate in the high-temperature limit. 

It is worth noting that the approach of $A^{(n\neq 0)}_E$ to unity at large $y$ is rather slow due to the first factor in \cref{Anhigher}. Nevertheless, $A^{(n\neq 0)}_E$ is strictly unity in the classical limit since the latter corresponds to strictly infinite $y$. To see this, let us restore the Planck constant $\hbar$. The quantities held fixed in the classical limit are the temperature $T$ and the field frequency $\Omega_m \equiv m/\hbar$. Thus, $y=\hbar^{-1}(2\pi T/\Omega_m)$ diverges when $\hbar$ is set to zero.

\subsection{Integration over zero mode}

The determinant in the $n=0$ sector vanishes because the critical bubble has a translational zero mode
\be
\label{zeromode}
\vf_{0,1}(x)=\sqrt{\frac{3\l T}{4m^3}}\, \phi_b^{\prime}(x)=-
\sqrt{\frac{3Tm}{2}}\cdot\frac{\sh{mx}}{\ch^2{mx}}\;,
\ee
where the coefficient has been fixed by the normalization requirement,
\be
\label{zeronorm}
\int \diff\tau \diff x\, \vf_{0,1}^2=1\;.
\ee
The zero mode satisfies the equation ${\cal O}_m \vf_{0,1}(x)=0$, and for $\mu=m$ the determinant ratio \eqref{fracdet} vanishes.
To regularize the ratio, we take $\mu$ slightly bigger than $m$. 
The corresponding operator ${\cal O}_\mu$ has a small eigenvalue $\a_{0,1}\approx \mu^2-m^2$. 
We divide out the contribution of this eigenvalue and obtain
\be
\label{zerofactor2}
\frac{\D}{\D^{(0)}}\mapsto 
\frac{\D^{\prime}}{\D^{(0)}}=\lim_{\mu\to m}
\frac{2\pi}{\mu^2-m^2}\frac{\D_\mu}{\D_{\mu}^{(0)}}=-\frac{2\pi}{12m^2}\;.
\ee
Note that we have included a factor of $2\pi$ because each mode in the Gaussian integration brings a factor $\sqrt{2\pi/\alpha_I}$ where $\alpha_I$ is the corresponding eigenvalue.

At the same time, we have to integrate over positions of the critical bubble.
To this end, we introduce unity into the original path integral (\ref{Zgen}):
\be
\label{shiftunity}
1=\int \diff b\, \delta\left(\int \diff\tau \diff x\, \phi(x,\tau) \vf_{0,1}(x+b) \right) \,
\left| \int \diff\tau \diff x\, \phi(x,\tau) \vf_{0,1}^{\prime}(x+b) \right|\;.
\ee
We then interchange the order of integration by taking the integral over 
$\diff b$ outside. The inner path integral is now constrained to the field configurations 
orthogonal to $\vf_{0,1}(x+b)$.
The saddle point of this integral is given by the shifted bubble $\phi_b(x+b)$. 
Due to the shift symmetry, the path integrals are identical for all $b$, hence the outer integral over $b$ just gives the total length $L$. 
In addition, we obtain a factor 
\be
\label{prefzero}
\left|\int \diff\tau \diff x\, \phi_b(x) \vf_{0,1}^{\prime}(x) \right|=
\sqrt{\frac{4m^3}{3\l T}}\;.
\ee   
Note that the expression inside the square root coincides with the
critical bubble action $S[\phi_b]$, as it should be~\cite{Coleman:1978ae}.
Combining \cref{zerofactor2,prefzero} into \cref{ImF} we obtain, 
\be
\label{ImF2}
\Im F=TL\cdot 2\sqrt{3}m\sqrt{\frac{E_b}{2\pi T}} \,A_E^{(n\neq 0)}\, \e^{-E_b/T}\;,
\ee
where $A_E^{(n\neq 0)}$ is given by \cref{Anhigher}.
Further using \cref{GammaAff,omminus} we arrive at the Euclidean prediction for the decay rate,
\be 
\label{GammaE}
\Gamma _E = \frac{6m^2}{\pi}\sqrt{\frac{E_b}{2\pi T}}\, A_E^{(n\neq 0)}\, \e^{-E_b/T} \;.
\ee
In the classical limit $A_E^{(n\neq 0)}=1$, and we obtain the prefactor \eqref{G_E2}.

\section{Numerical methods}
\label{app:num}

\subsection{Operator-splitting scheme for the Hamiltonian dynamics}

In what follows, we use the dimensionless units introduced in Sec.~\ref{ssec:units}. 
To solve numerically the equations of motion (\ref{Hameqs}), we use a pseudo-spectral, operator-splitting scheme~\cite{yoshida1990construction,mclachlan1993explicit}, which we presently review.
For convenience, we reproduce the equations here (for the sake of presentation, we suppress the spatial and wavenumber indices):
\be \label{Eqs1}
\left\lbrace \begin{array}{l}
\dot{\phi} = \pi  \\
\dot{\pi} = \Delta\phi-\phi-s\phi^3 \;,
\end{array}\right.
\ee
where we leave the sign $s=\pm 1$ of the self-interaction unspecified.
If $(\phi^t, \pi^t)$ is the solution of (\ref{Eqs1}) at time $t$, then at time $t+h$ the solution can be written as 
\be \label{EqSol1}
\left( \begin{array}{l} \phi^{t+h} \\ \pi^{t+h} \end{array} \right) = \e^{h{\cal O}} \left( \begin{array}{l} \phi^{t} \\ \pi^{t} \end{array} \right) \;.
\ee
We split the operator ${\cal O}$ as ${\cal O}={\cal L}+{\cal N}$, where 
\be \label{EqsNL1}
{\cal L}=\left( \begin{array}{cc}
0 & 1 \\ \Delta-1 & 0
\end{array} \right) \;, \quad {\cal N} = \left( \begin{array}{cc}
0 &0 \\ -s(\phi^t)^2 & 0
\end{array} \right)
\ee 
are the linear and non-linear parts of ${\cal O}$, respectively.
Note that for ${\cal O}={\cal N}$ or ${\cal O}={\cal L}$ the solution $(\phi^{t+h}, \pi^{t+h})$ can be written exactly.
Indeed, from \cref{EqsNL1} it follows that $\e^{h{\cal N}}=1+h{\cal N}$, hence the solution at time $t+h$ generated by ${\cal N}$ is 
\be \label{SolN1}
{\cal N}: \quad \phi^{t+h}=\phi^t\;, \quad \pi^{t+h} = \pi^t-hs(\phi^t)^3 \;.
\ee
To find the solution generated by ${\cal L}$, we switch to the Fourier space using \cref{IFT} and obtain the equations for the harmonic oscillator, whose solution at time $t+h$ is
\be \label{SolL1}
{\cal L}: \quad \begin{array}{l}
\tilde{\phi}^{t+h} = \tilde{\phi}^t\cos(h\O) + \tilde{\pi}^t \, \O^{-1}\sin(h\O) \;, \\
\tilde{\pi}^{t+h} = \tilde{\pi}^t\cos(h\O) - \tilde{\phi}^t \, \O\sin(h\O) \;,
\end{array}
\ee
where $\O^2=1+ 2(1-\cos ka)/a^2$ are the lattice mode frequencies.\footnote{Note that here the mode frequencies $\Omega^2$ do not contain the thermal mass correction, cf. \cref{Oj}.}

The idea of the splitting method is to approximate the evolution operator $\e^{\cal O}$ by a product of operators $\e^{\cal N}$ and $\e^{\cal L}$ whose action is known exactly.
We use the 4th order splitting:
\be \label{Split4}
\e^{h{\cal O}} = \prod_{i=4}^1 (\e^{h a_i {\cal L}}\e^{h b_i {\cal N}}) \;.
\ee
The coefficients $a_i$, $b_i$ can be found e.g. in~\cite{mclachlan1993explicit}.
Each exponential in this product acts according to \cref{SolN1,SolL1} (with the fractional time steps $a_kh$, $b_kh$) on the result of the previous steps.

\begin{figure}[t]
    \begin{minipage}[h]{0.49\linewidth}
        \centering
        \includegraphics[width=1.0\linewidth]{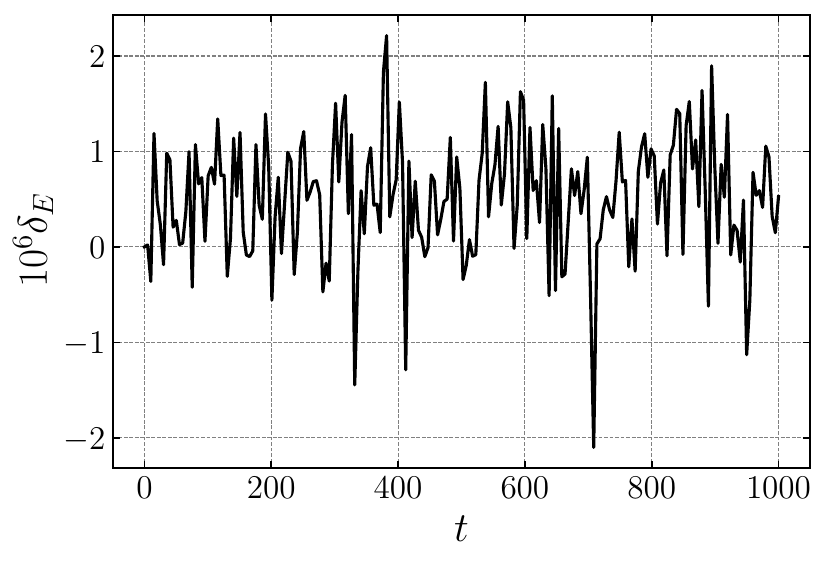} \\ (a)
	\end{minipage}
    \begin{minipage}[h]{0.49\linewidth}
        \centering
        \includegraphics[width=1.0\linewidth]{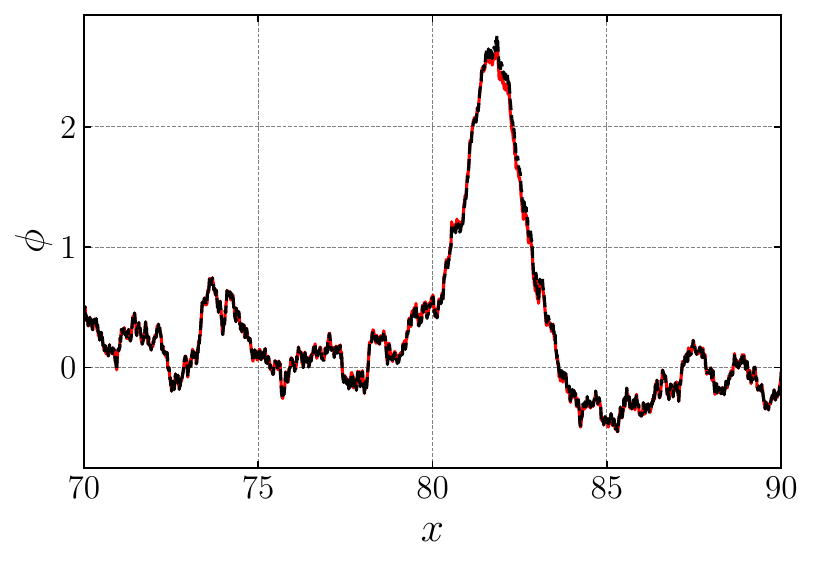} \\ (b)
	\end{minipage}
    \caption{\textbf{(a)} Relative energy error in a Hamiltonian simulation with positive self-interaction 
    ($s=+1$) and the parameters: $\TT=0.1$, $L=100$, $N=8192$, $h=10^{-2}$. \textbf{(b)} Two field configurations with negative self-coupling ($s=-1$) evolved from the same initial state with $h=10^{-2}$ (black dashed) and  $h=5\cdot 10^{-3}$ (red solid) at the moment of decay at $t=2473.14$. Other parameters are the same as in (a).}
	\label{fig:numtest1}
\end{figure}

The splitting (\ref{Split4}) breaks the energy conservation. 
In choosing the time step, we make sure that the relative energy error is negligible.
We constrain it to be no more than $\sim 10^{-6}$ throughout the simulation runs. 
Despite this stringent requirement, numerical tests show that the time step can be chosen rather large. It is sufficient to take $h/a\simeq 0.82$ to reach the desired precision, see Fig.~\ref{fig:numtest1}(a) for an illustration. The reason is that short modes with high frequencies, which usually put strongest constraints on the time step, are weakly interacting in the theory at hand. As such, they are almost linear and are evolved nearly exactly by our scheme.  

Next, we check that changing the time step by a factor of 2 does not affect the evolution of the field and, in particular, the decay time.
As an example, Fig.~\ref{fig:numtest1}(b) shows two decaying field configurations computed with $h=10^{-2}$ and $h=5\cdot 10^{-3}$ starting from the same initial configuration.
We see that the field profiles match each other and that the decay happens at the same time.

\subsection{Box size and lattice spacing}

Let us discuss the corrections due to the lattice size $L$ and spacing $a$. Since the critical bubble profile decreases exponentially with distance, the finite size effects on its shape and energy scale as $\e^{-L}$ and are negligible. Another important quantity is the thermal mass of the field entering the initial conditions (\ref{RJ}), (\ref{Oj}). The finite-size correction to it can be estimated as follows. In a finite box the momentum integral in the expression for the effective potential (\ref{V-1-loop-int}) gets replaced by a sum over discrete values of $p$. The main difference between the integral and the sum is at $p\sim 2\pi/L$. The corresponding effect on the thermal mass is
\be 
\delta \hat m_{\text{th}, L}^2 \sim \frac{\TT}{L} \;.
\ee
It is also negligible at $\TT\sim 0.1$ as soon as $L\gtrsim 10$.

\begin{figure}[t]
	\center{	
		\begin{minipage}[h]{0.49\linewidth}
			\center{\includegraphics[width=1.0\linewidth]{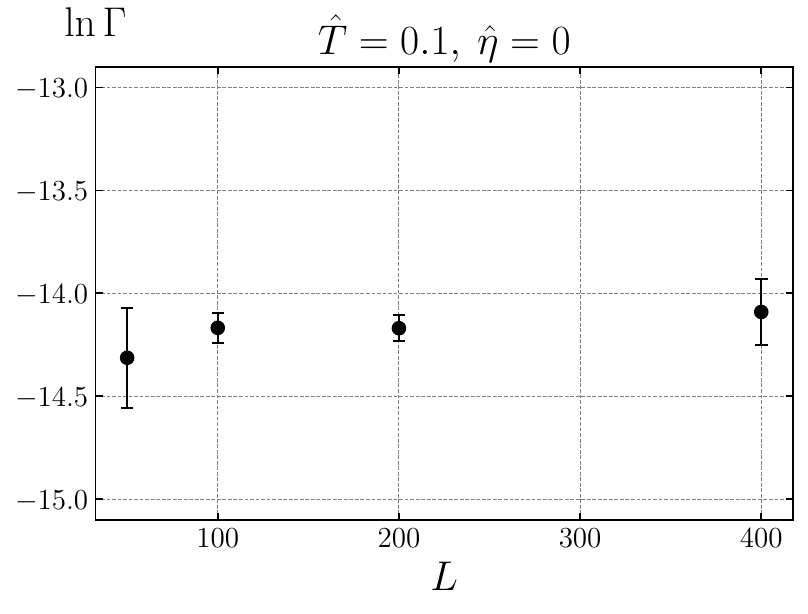}} \\ (a)
		\end{minipage}
		\begin{minipage}[h]{0.49\linewidth}
			\center{\includegraphics[width=1.0\linewidth]{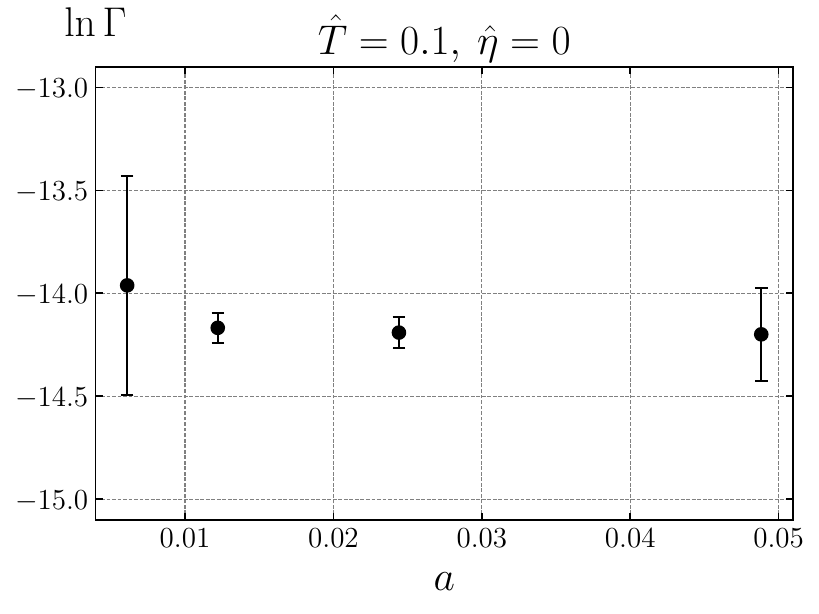}} \\ (b)
		\end{minipage}
	}
 \caption{Dependence of the measured decay rate in the Hamiltonian simulations on 
 \textbf{(a)} the lattice size $L$ at fixed spacing $a=1.2\cdot 10^{-2}$ and  
 \textbf{(b)} the spacing $a$ at fixed size $L=100$. The ratio of the time step to the lattice spacing 
  $h/a=0.82$ is fixed in both cases. Error bars show the statistical uncertainty of the measurements.}
	\label{fig:lat_tests}
\end{figure}

Further, we require that the energy of the subsystem consisting of long modes with wavenumbers $|k|\leq 2$ significantly exceeds the energy of the critical bubble. This is needed to avoid hindering the bubble nucleation by a shortage of energy in the relevant modes. Since each mode carries energy $\TT$, the total energy of long modes is $\hat E_{\rm long}\simeq 2L\TT/\pi$. Thus, $\hat E_{\rm long}\gg \Es$ implies $L\gg 20$. We have found that $L=100$ is sufficient. In Fig.~\ref{fig:lat_tests}(a) we show the values of the decay rate measured in the simulations of the Hamiltonian dynamics in the boxes of different size at fixed fiducial lattice spacing $a=1.2\cdot 10^{-2}$. We see that the results are independent of $L$ for $L\geq 100$. The point at $L=50$ appears to be slightly lower than the rest of the points, but still within the statistical uncertainty of the measurements. We conclude that the systematic error $\s^{\rm sys}_{\ln\G}$ in the determination of $\ln \Gamma$ does not exceed the statistical uncertainty $\s^{\rm stat}_{\ln\G}$. We conservatively take $\s^{\rm sys}_{\ln\G}=\s^{\rm stat}_{\ln\G}\sim 0.1$, where $\s^{\rm stat}_{\ln\G}$ is taken from the runs with $L=100$ and $L=200$ in which we acquired the largest statistics.

We now turn to the lattice spacing. Since we use the second-order discretization scheme (see \cref{E_d}), the corrections to the bubble shape and energy are $O(a^2)$. They are small at $a\sim 10^{-2}$. The same applies to the thermal mass. The latter can in fact be calculated exactly using the 1-loop lattice effective potential. Keeping only the contribution of the zero Matsubara mode we have (cf. \cref{V-1-loop-int}),
\be 
\label{Vefflat}
V_{\rm eff}(\bar{\phi}) =V(\bar\phi)+\frac{T}{2}\limitint_{-\pi/a}^{\pi/a}\frac{\diff p}{2\pi}\ln\Omega^2(p,\bar\phi) \;,
\ee
where $\Omega^2(p,\bar\phi)=2(1-\cos ap)/a^2+1-3\bar\phi^2$.
Expanding the integrand to quadratic order in $\bar{\phi}$ and evaluating the integral over momentum, we obtain
\be \label{mth_lat}
\hat m^2_{\rm th} = 1 - \frac{3\TT}{\sqrt{4+ a^2}} \;.
\ee
The $a$-dependent contribution is $O(\TT a^2)$ and is negligible. 

Finally, we directly explore the dependence of the decay rate on $a$. Fig.~\ref{fig:lat_tests}(b) shows $\ln \G$ measured in ensembles of simulations with different $a$ and fixed box size $L=100$. We see that the results are consistent with each other within the statistical uncertainty. The latter is $\s_{\ln\G}^{\rm stat}\sim0.1$ for the two sets of simulations at $a=1.2\cdot 10^{-2}$ and $a=2.4\cdot 10^{-2}$, and thus we conclude that the systematic error is below this value. 

\subsection{Operator-splitting scheme for the Langevin dynamics}

Here we outline the operator-splitting method to solve the Langevin equation (\ref{LangEq}).
We write it again as a system of first-order equations
\be \label{Eqs2}
\left\lbrace \begin{array}{l}
\dot{\phi} = \pi \;, \\
\dot{\pi} = \Delta\phi-\phi-s\phi^3 - \heta\pi + \s\hat{\xi} \;,
\end{array}\right.
\ee
with $\s^2=2\heta \TT$. 
If the force $\hat\xi(t)$ were a smooth function of time, we could proceed as before and write the solution between $t$ and $t+h$ as in \cref{EqSol1}, where ${\cal O}$ is the operator associated with the system (\ref{Eqs2}). The force 
 and linear dissipation terms in (\ref{Eqs2}) would belong to the linear part of ${\cal O}$, and the evolution due to the linear part could still be solved exactly.

Complications arise because $\hat\xi(t)$ is not smooth: it is a white noise having arbitrarily sharp variations. Developing accurate numerical scheme for this case is non-trivial~\cite{kloeden2013numerical}. The idea is to replace eqs.~(\ref{Eqs2}) on every time step by another system with a regular force term in such a way that the solutions of the two systems have the same statistical properties, up to the required order of precision. The simplest choice would be replacing 
$\hat{\xi}(t,x)$ by a sequence of independent Gaussian random variables $\xi^t_i$ with zero mean and variance
\be 
\ev{\xi^t_i \xi^t_j} = \frac{1}{h}\cdot \frac{\delta_{ij}}{a} \;,
\ee
where $i,j$ label the lattice sites. 
However, such treatment of the stochastic term reduces the order of convergence to at most 1 \cite{telatovich2017strong}. 
To achieve better convergence, one needs to introduce additional random variables entering the r.h.s. of the equation in a specific way~\cite{kunita1980representation, misawa2000numerical}. 

\subsubsection{3rd order scheme}

Let $(\phi^t, \pi^t)$ be the solution of (\ref{Eqs2}) at time $t$. 
Following~\cite{telatovich2017strong}, we write the solution at time $t+h$ as
\be \label{EqSol3}
\left( \begin{array}{l} \phi^{t+h} \\ \pi^{t+h} \end{array} \right) = \e^{h{\cal O}_3} \left( \begin{array}{l} \phi^{t} \\ \pi^{t} \end{array} \right) \;,
\ee
where the operator ${\cal O}_3$ is associated with the following system of equations to be solved at $(t,t+h)$:
\be \label{Eqs4}
\left\lbrace \begin{array}{l}
\dot{\phi} = \pi + \s\xi_2^t - \heta\s\xi_3^t \;,\\
\dot{\pi} = \Delta\phi-\phi-s\phi^3 -\heta\pi+ \s\xi_1^t - \s\heta\xi_2^t + \s\heta^2\xi_3^t + \s\Delta\xi_3^t - \s\xi_3^t - 3s\phi^2\s\xi_3^t \;.
\end{array}\right.
\ee
Here $\xi_1^t$, $\xi_2^t$ and $\xi_3^t$ are Gaussian random variables with zero mean and variance
\be 
    \ev{\xi^t_{b,i} \xi^t_{c,j}} = C_{3,bc}\cdot\frac{\delta_{ij}}{a} \;,
\ee
where the correlation matrix is given by\footnote{This correlation matrix differs from the one used in \cite{telatovich2017strong} at the third order. We believe the discrepancy is due to a typo that can be traced back to \cite{kunita1980representation}.}
\be 
C_3 = \left( \begin{array}{ccc}
\frac{1}{h} & 0 & 0 \\
0 & \frac{h}{12} & 0 \\
0 & 0 & \frac{h^3}{720}
\end{array} \right) \;.
\ee
The operator ${\cal O}_3$ leads to the 3rd order convergence (see below) as soon as the order of the subsequent splitting is not smaller than 3.

We can now proceed as above and split the evolution operator into the linear and non-linear parts,
${\cal O}_3={\cal L}_3+{\cal N}_3$. 
The evolution due to each of these parts is known exactly.
The solution at $t+h$ generated by ${\cal N}_3$ is given by
\be \label{SolN2}
{\cal N}_3: \quad \phi^{t+h}=\phi^t\;, \quad \pi^{t+h} = \pi^t-hs(\phi^t)^3 - 3hs(\phi^t)^2\s\xi_3^t \;.
\ee
The solution generated by ${\cal L}_3$ in Fourier space satisfies the equations for the damped harmonic oscillator in the presence of external force:
\be \label{EqsL2}
{\cal L}_3: \quad \left\lbrace\begin{array}{l}
\dot{\tilde{\phi}} = \tilde{\pi} + \s\tilde{\mu}_1 \;, \\
\dot{\tilde{\pi}} = -\O^2\tilde{\phi} - \heta\tilde{\pi} + \s\tilde{\mu}_2 \;,
\end{array}\right.
\ee
where we denoted 
\be 
\tilde\mu_1=\tilde\xi_2^t - \heta\tilde\xi_3^t \;, \qquad \tilde\mu_2 = \tilde\xi_1^t - \heta\tilde\xi_2^t+\heta^2\tilde\xi_3^t-\Omega^2\tilde\xi_3^t \;.
\ee
For the under-damped modes, the solution to these equations at time $t+h$ takes the form,
\be \label{SolL2}
{\cal L}_3: \quad \begin{array}{l}
\tilde{\phi}^{t+h} = \left( a_{\phi} \cos(h\O_\eta) + b_{\phi}\sin(h\O_\eta) \right) \e^{-\frac{\heta h}{2}} + c_{\phi} \;, \\
\tilde{\pi}^{t+h} = \left( a_{\pi} \cos(h\O_\eta) + b_{\pi}\sin(h\O_\eta) \right) \e^{-\frac{\heta h}{2}} + c_{\pi} \;,
\end{array}
\ee
where $\O_\eta = \sqrt{\O^2-\heta^2/4}$ and
\be
\begin{aligned}
& a_{\phi} = \tilde{\phi}^t - \frac{\s(\heta\tilde{\mu}_1+ \tilde{\mu}_2)}{\O^2} \;, \\
& b_{\phi} = \frac{1}{\O_\eta}\left( \tilde{\pi}^t + \frac{\heta\tilde{\phi}^t}{2} \right) + \frac{\s}{\O_\eta}\left( \tilde{\mu}_1 - \frac{\heta (\heta\tilde{\mu}_1+ \tilde{\mu}_2)}{2\O^2} \right) \;, \\
& c_{\phi} = \frac{\s(\heta\tilde{\mu}_1+ \tilde{\mu}_2)}{\O^2} \;, \\
& a_{\pi} = \tilde{\pi}^t + \s\tilde{\mu}_1 \;, \\
& b_{\pi} = -\frac{1}{\O_\eta}\left( \O^2\tilde{\phi}^t + \frac{\heta\tilde{\pi}^t}{2} \right) + \frac{\s (\heta\tilde{\mu}_1+2\tilde{\mu}_2)}{2\O_\eta} \;, \\
& c_{\pi} = -\s\tilde{\mu}_1  \;.
\end{aligned}
\ee
For the over-damped modes $\O_\eta = \sqrt{\heta^2/4-\O^2}$ and one should replace $\cos, \sin\mapsto\ch, \sh$ in the above expressions.
Finally, we split the operator exponent in (\ref{EqSol3}) as in \cref{Split4}. 
We use the 3rd order splitting~\cite{Ruth:1983ad}.\footnote{We have also tested the scheme with the 4th order splitting and found that it gives essentially the same precision.}
The action of each exponential in the product is then evaluated according to \cref{SolN2,SolL2} (with the appropriate fractional time steps).

\begin{figure}[t]
	\centering
    \includegraphics[width=0.6\linewidth]{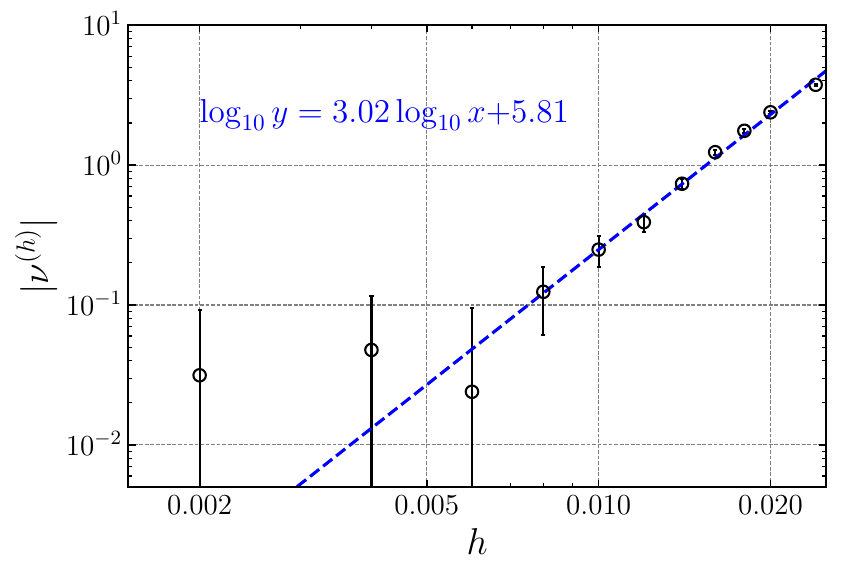}
    \caption{Measurement of the (weak) order of convergence of the operator-splitting, pseudo-spectral scheme for the Langevin equation (\ref{Eqs2}) ($s=+1$). Vertical axis shows the error in the momentum variance averaged over $M=100$ configurations, and the horizontal axis shows the time step. We use the stochastic 3rd order scheme (\ref{Eqs4}) with 3rd order operator splitting (cf. \cref{Split4}).
    The blue dashed line is the fit of the data points, its equation is given in the plot.
    The first and the last points are excluded from the fit.
    We take $\TT=0.1$, $\hat{\eta}=1$, and $t_f=10$. }
    \label{fig:conv_weak_33}
\end{figure}

We have directly tested the convergence of the scheme as follows.
We fix the value of $h$ and evolve a set of $M$ configurations starting with the same initial configuration but with different realizations of the random variables $\xi^t_1$, $\xi_2^t$, $\xi_3^t$ (or $\mu_1$, $\mu_2$, $\mu_3$ at each time step).
The initial configuration is prepared according to the thermal distribution (\ref{RJ}).
At the end of simulation, $t=t_f\gg h$, we measure the mean error of the momentum variance across the Fourier modes for each configuration,
\be \label{weak_1}
\nu_J^{(h)} = \frac{1}{N} \sum_{i=0}^{N-1} \left| \tilde{\pi}_i^{(h,J)} \right|^2 - \frac{\TT}{a} \;, \qquad J=1, \dots ,M \;.
\ee
The mean error and the variance of this mean are
\be \label{weak_2}
\nu^{(h)} = \frac{1}{M} \sum_{J=1}^M \nu_J^{(h)} \;, \qquad \s_{\nu^{(h)}}^2 = \frac{1}{M^2}\sum_{J=1}^M \left(\nu_J^{(h)}-\nu^{(h)}\right)^2 \;.
\ee
We repeat the exercise for different values of $h$ and fit the data points by a straight line in the log-log plot. 
The result is shown in Fig.~\ref{fig:conv_weak_33}.
Note that for a fixed set size $M$ the variance has non-vanishing value $\sigma_{\nu^{(0)}}$
in the limit $h\to 0$. The numerical error due to finite $h$ exceeds $\sigma_{\nu^{(0)}}$ at $h\gtrsim 0.01$ but drops below this threshold at smaller $h$. We see that at $h\gtrsim 0.01$ 
the error $\nu^{(h)}$ 
behaves as
\be \label{weak_3}
\nu^{(h)}\simeq\const\cdot  h^{3.02}\;,
\ee
which is consistent with the expected behavior $\nu^{(h)}\propto h^3$.

\begin{figure}[t]
    \begin{minipage}[h]{0.49\linewidth}
        \centering
        \includegraphics[width=1.1\linewidth]{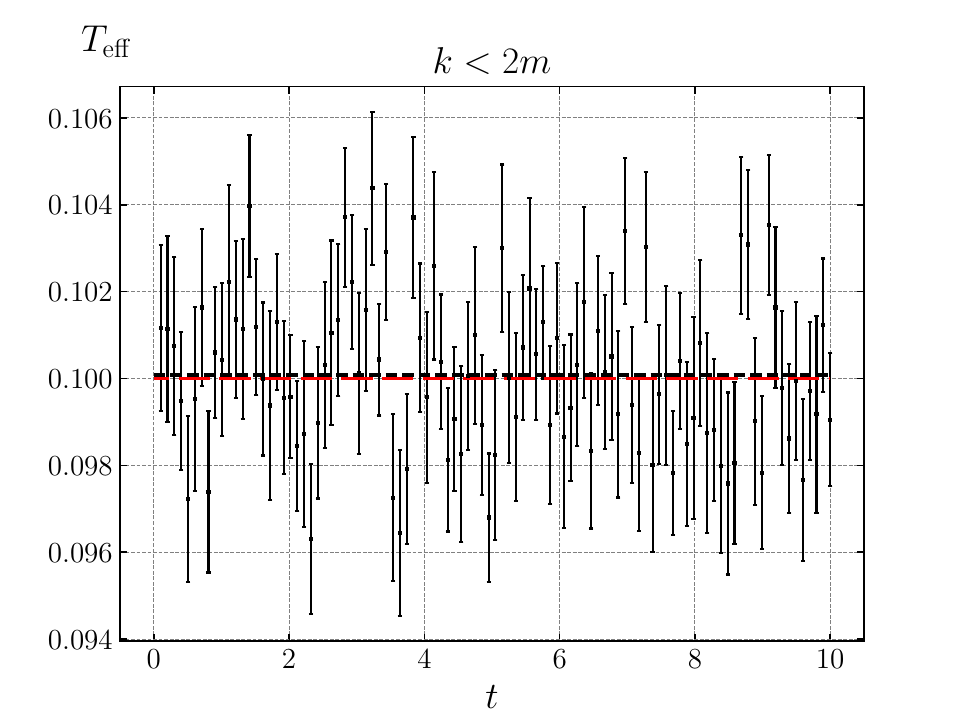} \\ (a)
	\end{minipage}
	\begin{minipage}[h]{0.49\linewidth}
        \centering
        \includegraphics[width=1.1\linewidth]{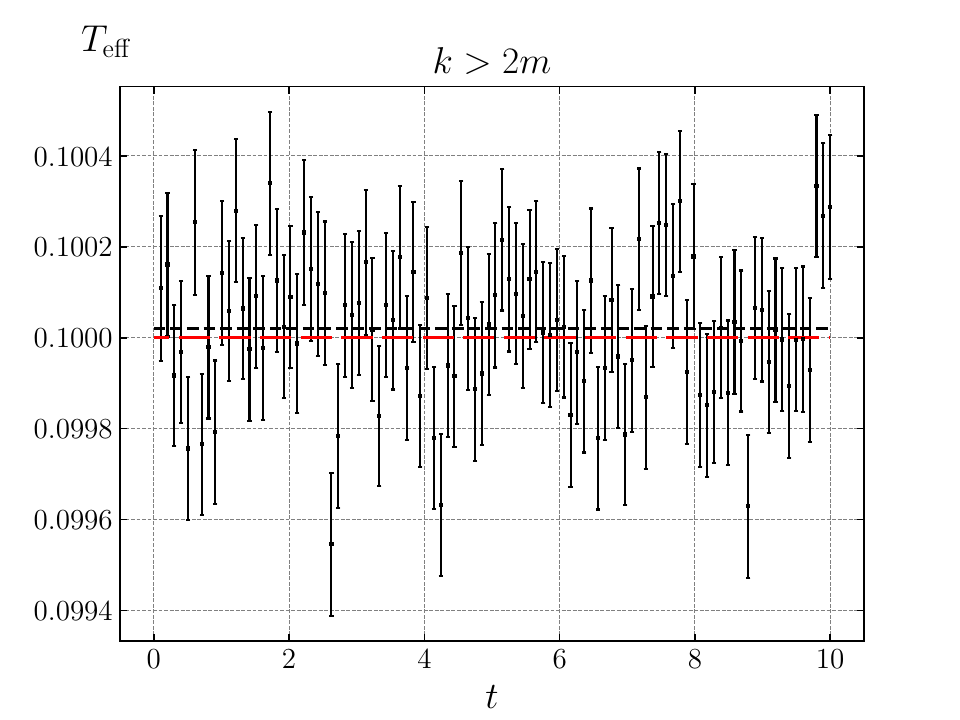} \\ (b)
	\end{minipage}
    \caption{Measurement of the effective temperature (\ref{Teff}) of long \textbf{(a)} and short \textbf{(b)} modes during the Langevin evolution (\ref{Eqs2}) ($s=+1$). The field is evolved using the operator-splitting, pseudo-spectral scheme with $h=2.5\cdot 10^{-3}$. The error bars represent the statistical uncertainty. The average is over 100 simulations with different realizations of the thermal initial state (\ref{RJ}) and noise. The black dashed line denotes the time-averaged value of $T_{\rm eff}$, and the red dash-dotted line denotes the theoretical value $\TT=0.1$. We take $\heta=1$.}
	\label{fig:app:Teff}
\end{figure}

To determine the optimal time step, we measure the 
statistical properties of the ensemble and make sure that they are preserved in time.
Two important observables are the effective temperatures (\ref{Teff}) of long (relevant for the decay) and short modes.
We require the systematic error of both observables, introduced by the finite time step, not to exceed~$0.1\%$.
Such accuracy in determining the temperature of the long modes translates into the $\sim 1\%$ estimate for the numerical error in the measurement of the prefactor, which is well below the statistical uncertainty (see Fig.~\ref{fig:pref_Eta_temp}).

Numerical tests show that it is sufficient to take the time step $h=2.5\cdot 10^{-3}$ and the lattice spacing $a=1.2\cdot 10^{-2}$.
Fig.~\ref{fig:app:Teff} shows the measured values of $T_{\rm eff}$ during the field evolution with a large dissipation coefficient $\heta=1$, averaged over 100 realizations.
The realizations are prepared in the thermal state with $\TT=0.1$, and we see that, on average, $T_{\rm eff}=\TT$ with the required precision.

\section{Statistical error in the measurement of the decay rate}
\label{app:stat}

Consider an ensemble of $N_0$ identical systems that decay according to an exponential law, so that the expectation value of  the number of survivors after time $t$ is 
\be
\label{avsurv}
\ev{N(t)} =N_0\e^{-\Gamma t}\;.
\ee
The actual number of survivors fluctuates, and the rate inferred from it has a statistical error. 
Here we derive an estimate for this error.

Let us introduce the measured survival probability $P_{surv}(t)=N(t)/N_0$ and consider
the following estimator for the decay rate:
\be
\label{estim}
\hat{\Gamma}=-\frac{1}{t}\ln P_{surv}(t)\;.
\ee
Its expectation value and variance are given by
\be
\label{avdisp}
\langle\hat{\Gamma}\rangle = -\frac{1}{t}\big( \ev{\ln N(t)} -
\ln N_0 \big) \;, \qquad
\sigma_\Gamma^2=\frac{1}{t^2} \left( \ev{(\ln N(t))^2} -
\ev{\ln N(t)}^2 \right) \;.
\ee
To compute these we need the distribution of $N(t)$. For a single system the probability to survive is $p_{\rm s}=\e^{-\Gamma t}$, and the probability to decay is $p_{\rm d}=1-\e^{-\Gamma t}$. 
The probability $p(N)$ that exactly $N$ systems have survived after time
$t$ (and hence $N_0-N$ systems decayed) is given by the binomial distribution, $p(N)=C_{N_0}^N \; p_{\rm s}^{N}\, p_{\rm d}^{N_0-N}$.
When the size of the ensemble is large, $N_0, N, (N_0-N) \gg 1$, the latter is well approximated by the Gaussian distribution,
\be
\label{gauss}
p(N)=\frac{1}{\sqrt{2\pi N_0 p_{\rm s}(1-p_{\rm s})}}
\exp\left[-\frac{(N-p_{\rm s}N_0)^2}{2N_0 p_{\rm s}(1-p_{\rm s})}\right]\;,
\ee 
which indeed corresponds to the average number of survivors (\ref{avsurv}), $\ev{N} =N_0p_{\rm s}$. The variance is $\sigma_{N}^2=N_0 p_{\rm s}(1-p_{\rm s})$.
Using the distribution (\ref{gauss}) we compute the average 
of the decay width estimator,
\be
\label{avestim}
\langle{\hat{\Gamma}}\rangle = \Gamma+ \frac{\e^{\Gamma t}-1}{2N_0 t}\;,
\ee 
where we have neglected terms suppressed by additional powers of $\sigma^2_N/\ev{N}^2$. 
We see that the estimator is slightly biased, but the bias is small as long as the number of survivors remains large, $N_0\e^{-\Gamma t}\gg 1$. 
Next, we compute the average $\ev{(\ln N)^2}$ and get for the variance:
\be \label{Gammavar}
\sigma_\Gamma^2=\frac{\e^{\Gamma t}-1}{N_0 t^2}\;.
\ee
Let us focus on the case when the fraction of decayed systems is small, $\Gamma t\ll 1$, but their number $N_{\rm decay}$ is still large.
Then, we have
\be
\label{reldisp}
\sigma_\Gamma^2\approx \frac{\Gamma}{N_0 t} \quad \Longrightarrow \quad
\frac{\sigma_\Gamma}{\Gamma}=\frac{1}{\sqrt{N_0 \Gamma
    t}}=\frac{1}{\sqrt{N_{\rm decay}(t)}}\;.
\ee
We obtain that the relative statistical uncertainty in the determination of the decay rate over a certain time interval $t$ is the inverse square root of the number of decays during this interval. 
This agrees with the naive Poissonian estimate:
\be
\Gamma\simeq \frac{N_{\rm decay}(t)}{t}\quad\Longrightarrow\quad
\frac{\s_\G}{\G}\sim \frac{\s_{N_{\rm decay}}}{N_{\rm decay}}\sim\frac{1}{\sqrt{N_{\rm decay}}}\;.
\ee
For $N_{\rm decay}$ fixed, the absolute error will be larger for larger rate. 

\bibliographystyle{JHEP}
\bibliography{Refs.bib}

\end{document}